\documentclass{llncs}

\usepackage{latexsym,amssymb,amsmath}
\usepackage{algorithm}
\usepackage[noend]{algpseudocode}
\usepackage{xspace}
\usepackage{url}
\usepackage{enumerate}
\usepackage[defblank]{paralist}
\usepackage{graphicx}
\graphicspath{{imgs/}}
\usepackage{todonotes}
\usepackage[defblank]{paralist}
 \usepackage{comment}
 \usepackage{todonotes}
 \usepackage{hyperref}
 \usepackage{tabularx}
 \usepackage{booktabs}
 \usepackage{multirow}
  \usepackage{sidecap}

\allowdisplaybreaks[2]

\usepackage{times}

\usepackage{tikz}
\newcommand*\circled[1]{\tikz[baseline=(char.base)]{
    \node[shape=circle,draw,inner sep=1pt] (char) {#1};}}

\newcommand{\Nat}{\ensuremath{\mathrm{I\kern-1.5pt N}}}

\def\mi#1{\mathit{#1}}

\def\post#1{\ensuremath{{#1}\kern-.05ex\bullet}\,}

\def\postnet#1#2{\ensuremath{{#2}\!\kern-.05ex\stackrel{#1}{\bullet}}\,}

\def\posto#1{\ensuremath{{#1}\kern-.05ex\circ}\,}

\def\postonet#1#2{\ensuremath{{#2}\!\kern-.05ex\stackrel{#1}{\circ}}\,}

\def\Nat{\ensuremath{\mathrm{I\kern-1.5pt N}}}

\newcommand{\ebefore}[2]{\mkern-2mu\lhd_{#1}\mkern-2.5mu(#2)}
\newcommand{\eafter}[2]{\mkern-2mu\rhd_{#1}\mkern-2.5mu(#2)}
\newcommand{\eincbefore}[2]{\mkern-2mu\unlhd_{#1}\mkern-2.5mu(#2)}
\newcommand{\eincafter}[2]{\mkern-2mu\unrhd_{#1}\mkern-2.5mu(#2)}

\newcommand{\alwys}{\Box}
\newcommand{\eventy}{\diamondsuit}

\newcommand{\cardsrcalways}{\sharp^{\alwys}_{\mi{src}}}
\newcommand{\cardsrceven}{\sharp^{\eventy}_{\mi{src}}}
\newcommand{\cardtaralways}{\sharp^{\alwys}_{\mi{tar}}}
\newcommand{\cardtareven}{\sharp^{\eventy}_{\mi{tar}}}

\newcommand{\acardalways}{\sharp^{\alwys}_{\mi{A}}}
\newcommand{\acardeven}{\sharp^{\eventy}_{\mi{A}}}
\newcommand{\ocard}{\sharp_{\mi{OC}}}

\makeatletter
\newcommand*{\inlineequation}[2][]{%
  \begingroup
    \refstepcounter{equation}%
    \ifx\\#1\\%
    \else
      \label{#1}%
    \fi
    \relpenalty=10000 %
    \binoppenalty=10000 %
    \ensuremath{%
      #2%
    }%
    ~\@eqnnum
  \endgroup
}
\makeatother

\newif\ifdraft
\drafttrue

\title{Object-Centric Behavioral Constraints}
\titlerunning{Object-Centric Behavioral Constraints}

\author{Wil M.P. van der Aalst\inst{1} \and Guangming Li\inst{1} \and Marco Montali\inst{2}}  %
\authorrunning{Wil van der Aalst, Guangming Li and Marco Montali}

\institute{Eindhoven University of Technology,
P.O.\ Box 513, 5600~MB, Eindhoven, The Netherlands. \email{w.m.p.v.d.aalst@tue.nl, g.li.3@tue.nl}
\and Free University of Bozen-Bolzano, Piazza Domenicani 3, I-39100, Bolzano, Italy. \email{montali@inf.unibz.it}}

\begin{document}

\maketitle
\sloppy

\begin{abstract}
Today's process modeling languages often force the analyst or modeler to straightjacket real-life processes into
simplistic or incomplete models that fail to capture the essential
features of the domain under study.
Conventional business process models only describe the lifecycles of individual instances (cases) in isolation.
Although process models may include data elements (cf.\ BPMN), explicit connections
to \emph{real} data models (e.g.\ an entity relationship model or a UML class model) are rarely made.
Therefore, we propose a novel approach that \emph{extends data models with a behavioral perspective}.
Data models can easily deal with many-to-many and one-to-many relationships.
This is exploited to create process models that can also model complex interactions between different types of instances.
Classical multiple-instance problems are circumvented by using the data model for event correlation.
The declarative nature of the proposed language makes it possible to
model behavioral constraints over activities like cardinality constraints in data models.
The resulting \emph{object-centric behavioral constraint (OCBC) model} is able to describe processes
involving \emph{interacting instances} and \emph{complex data dependencies}.
In this paper, we introduce the \emph{OCBC model and notation}, providing a number of examples that give a flavour of the approach. We then define a \emph{set-theoretic semantics} exploiting cardinality constraints within and across time points. We finally formalize \emph{conformance checking} in our setting, arguing that evaluating conformance against OCBC models requires diagnostics that go beyond what is provided by contemporary conformance checking approaches. 
\end{abstract}

\section{Introduction}\label{sec:intro}

Techniques for business process modeling (e.g., BPMN diagrams, Workflow nets, EPCs, or UML activity diagrams)
tend to suffer from two main problems:
\begin{compactitem}
  \item It is \emph{difficult to model interactions between process
      instances}, which are in fact typically considered in
    isolation. Concepts like lanes, pools, and message flows in
    conventional languages like BPMN aim to address this. However, within each (sub)process still a single instance is modeled in isolation.
  \item It is also \emph{difficult to model the data-perspective and control-flow perspective in a
      unified and integrated manner}. Data objects can be modeled, but the more powerful
    constructs present in Entity Relationship (ER) models and UML
    class models cannot be expressed well in process models. For example, cardinality
    constraints in the data model \emph{must} influence behavior, but
    this is not reflected at all in today's process models.
\end{compactitem}

Because of these problems there is a mismatch between process models and the data in (and functionality supported by) real enterprise systems from vendors such as SAP (S/4HANA), Microsoft (Dynamics 365), Oracle (E-Business Suite), and Salesforce (CRM). These systems are also known as Enterprise Resource Planning (ERP) and/or Customer Relationship Management (CRM) systems and support business functions related to sales, procurement, production, accounting, etc. These systems may contain hundreds, if not thousands, of tables with information about customers, orders, deliveries, etc.
For example, SAP has tens of thousands of tables.
Also Hospital Information Systems (HIS) and Product Lifecycle Management (PLM) systems have information about many different entities scattered over a surprising number of database tables.
Even though a clear process instance notion is missing in such systems,
mainstream business process modeling notations can only describe the lifecycle of one type of process instance at a time.
The disconnect between process models and the actual processes and systems
becomes clear when applying process mining using data from enterprise systems.
How to discover process models or check conformance if there is no single process instance notion?

The problems mentioned have been around for quite some time (see for example \cite{heeboek}),
but were never solved satisfactorily.
Artifact-centric approaches  \cite{CohnH:2009:business_artifacts,GSM-Hull-2011,compliance-artifacts-bpm2011,NigamC:2003:artifacts}
(including the earlier work on proclets \cite{aalprocletsjcis}) attempt to address the above problems.
However, these approaches tend to result in models where
\begin{compactitem}
  \item the description of the end-to-end behavior needs to be distributed over multiple diagrams (e.g., one process model per artifact),
  \item the control-flow cannot be related to an overall data model (i.e., there is no explicit data model or it is separated from the control-flow), and
  \item interactions between different entities are not visible or separated (because artifacts are distributed over multiple diagrams).
\end{compactitem}
Within an artifact, proclet, or subprocess, \emph{one is forced to pick a single instance notion}.
Moreover, cardinality constraints in the data model cannot be
exploited while specifying the intended dynamic behavior.
We believe that data and process perspectives can be unified better, as demonstrated in this paper.

\begin{SCfigure}
\centering
\includegraphics[width=5cm]{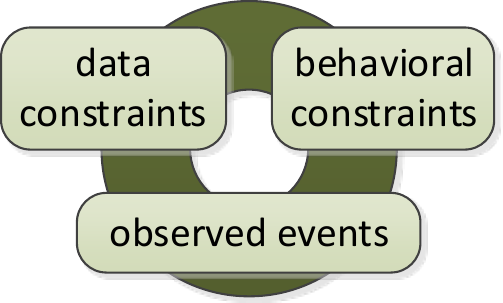}
\caption{Object-Centric Behavioral Constraint (OCBC) models connect data constraints (like in a UML class diagram), behavioral constraints (like in a process model or rule set), and real event data. This allows for novel forms of conformance checking.}\label{fig-overview}
\end{SCfigure}
This paper proposes the \emph{Object-Centric Behavioral Constraint} (OCBC) model
as a novel language that combines ideas from declarative, constraint-based languages like \emph{Declare} \cite{declareCSRD09},
and from data/object modeling techniques (ER, UML, or
ORM). Cardinality constrains are used as a \emph{unifying mechanism} to
tackle data and behavioral dependencies, as well as their interplay (cf.\ Figure~\ref{fig-overview}).
The primary application considered in this paper is \emph{conformance checking} \cite{wires-replay,BIS-artifactconformance-lnbip2011,BIS-data-aware-checking-lnbip2012,anne_confcheck_is}.
Deviations between observed behavior (i.e., an event log) and modeled behavior (OCBC model) are diagnosed
for compliance, auditing, or risk analysis.
Unlike existing approaches, instances are not considered in isolation and cardinality constraints in the data/object model are taken into account.
Hence, problems that would have remained undetected earlier, can now be detected.

\begin{figure}[t]
\begin{center}
\includegraphics[width=12cm]{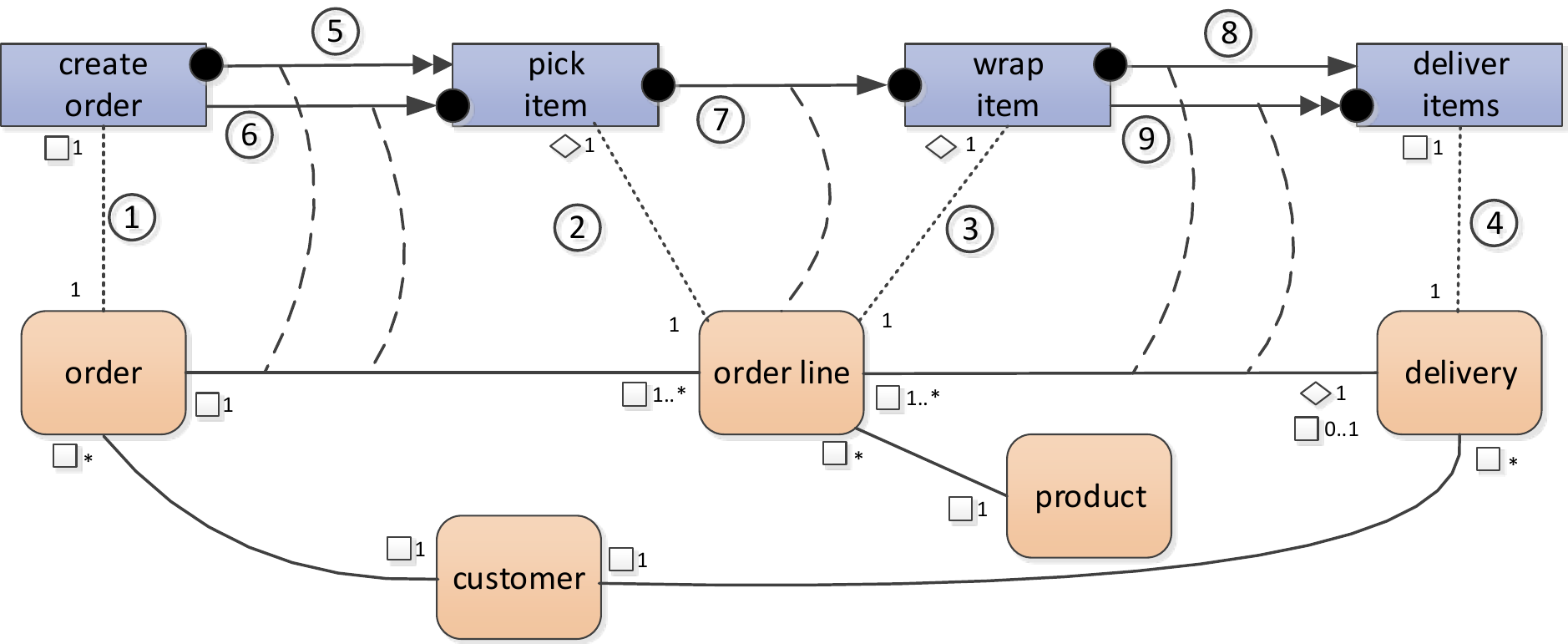}
\caption{A small \emph{Object-Centric Behavioral Constraint} (OCBC) model. }\label{fig-intro}
\end{center}
\end{figure}

Figure~\ref{fig-intro} shows an OCBC model with four activities (\emph{create order}, \emph{pick item}, \emph{wrap item}, and \emph{deliver items})
and five object classes (\emph{order}, \emph{order line}, \emph{delivery}, \emph{product}, and \emph{customer}).
The top part describes the ordering of activities and the bottom part the structuring of objects relevant for the process.
The lower part can be read as if it was a UML class diagram.
Some cardinality constraints should hold at any point in time as indicated by the $\alwys$ (``always'') symbol.
Other cardinality constraints should hold from some point onwards as indicated by the $\eventy$ (``eventually'') symbol.
Consider for example the relation between \emph{order line} and \emph{delivery}.
At any point in time a delivery corresponds to one of more order lines (denoted $\alwys~1..^\ast$)
and an order line refers to as most one delivery (denoted $\alwys~0..1$).
However, eventually an order line should refer to precisely one delivery (denoted $\eventy~1$).
Always, an order has one or more order lines, each order line corresponds to precisely one order,
each order line refers to one product, each order refers to one customer, etc. The top part shows behavioral constraints and the middle part relates
activities, constraints, and classes.

The notation used in Figure~\ref{fig-intro} will be explained in more detail later.
However, to introduce the main concepts, we first informally describe the 9 constructs highlighted in Figure~\ref{fig-intro}.
\circled{1} There is a one-to-one correspondence between \emph{order} objects and \emph{create order} activities.
If an object is added to the class \emph{order}, the corresponding activity needs to be executed and vice versa.
\circled{2} There is a one-to-one correspondence between \emph{order line} objects and \emph{pick item} activities.
\circled{3} There is also a bijection between \emph{order line} objects and \emph{wrap item} activities.
The $\eventy~1$ annotations next to \emph{pick item} and \emph{wrap item} indicate that these activities need to be executed
for every order line. However, they may be executed some time after the order order line is created.
\circled{4} There is a one-to-one correspondence between \emph{delivery} objects and \emph{delivery items} activities.
\circled{5} Each \emph{create order} activity is followed by one or more \emph{pick item} activities related to the same order.
\circled{6} Each \emph{pick item} activity is preceded by precisely one corresponding \emph{create order} activity.
\circled{7} Each \emph{pick item} activity is followed by one \emph{wrap item} activity corresponding to the same order line.
Each \emph{wrap item} activity is preceded by one \emph{pick item} activity corresponding to the same order line.
\circled{8} Each \emph{wrap item} activity is followed by precisely one corresponding \emph{deliver items} activity.
\circled{9} Each \emph{deliver items} activity is preceded by at least one corresponding \emph{wrap item} activity.
A \emph{deliver items} activity is implicitly related to a set of order lines through the relationship
between class \emph{order line} and class \emph{delivery}.
The notation will be explained later, however, note that a single order may have many order lines
that are scattered over multiple deliveries.
Moreover, one delivery may combine items from multiple orders for the same customer.

The process described in Figure~\ref{fig-intro} cannot be modeled using conventional notations (e.g., BPMN)
because (a) three different types of instances are intertwined and (b)
constraints in the class model influence the allowed behavior.
Moreover, the OCBC model provides a \emph{full} specification of
the allowed behavior in a \emph{single diagram}, so that no further coding or annotation is needed.

The contribution of this paper is threefold:
\begin{compactenum}
\item we introduce the \emph{OCBC model and notation}, providing a number of examples that give a flavour of the approach;
\item we define a \emph{set-theoretic semantics} for OCBC models, exploiting cardinality constraints within and across time points;
\item we formalize \emph{conformance checking} in our setting, arguing that evaluating conformance against OCBC models requires diagnostics that go beyond what is provided by contemporary conformance checking approaches.
\end{compactenum}

The remainder is organized as follows.
Section~\ref{sec:behavconstr} focuses on behavioral constraints between tasks, showing in particular that declarative control flow constraints a l\`a Declare can be modeled as cardinality constraints over sets of events (i.e., task occurrences).
Section~\ref{sec:dataconstr} focuses on data modeling, first focusing on the intentional level of classes and relationships, then tackling the extensional level of objects and tuples.
Both aspects are integrated in Section~\ref{sec:join-two-worlds}, where the OCBC model is introduced.
Section~\ref{sec:conf-check} provides the semantics for OCBC models in terms of cardinality constraints over sets of events, and demonstrates that novel forms of conformance checking are possible. Section~\ref{sec:concl} concludes the paper.

\section{Modeling Behavioral Cardinality Constraints}\label{sec:behavconstr}

In this paper, a process is merely a collection of \emph{events} without assuming some case or process instance notion.
Each event corresponds to an \emph{activity} and may have additional \emph{attributes} such as
the time at which the event took place,
the resource executing the corresponding event,
the type of event (e.g., start, complete, schedule, abort),
the location of the event, or
the costs of an event.
Each event attribute has a \emph{name} (e.g., ``age'') and a \emph{value} (e.g., ``49 years'').
Moreover, events are atomic and ordered. For simplicity, we assume a \emph{total order}.
To model the overlapping of activities in time one can use start and complete events.

\begin{definition}[Events and Activities]\label{def:events}
${\cal U}_{E}$ is the universe of \emph{events}, i.e., things that happened.
${\cal U}_{A}$ is the universe of \emph{activities}.
\begin{compactitem}
  \item Function $\mi{act} \in {\cal U}_{E} \rightarrow {\cal U}_{A}$ maps events onto activities.
  \item Events can also have additional attributes (e.g., timestamp, cost, resource, or amount).
${\cal U}_{\mi{Attr}}$ is the universe of \emph{attribute names}.
${\cal U}_{\mi{Val}}$ is the universe of \emph{attribute values}.
$\mi{attr} \in {\cal U}_{E} \rightarrow ( {\cal U}_{\mi{Attr}} \not\rightarrow {\cal U}_{\mi{Val}})$ maps events onto a partial function assigning values to some attributes,\footnote{$f \in X \not\rightarrow Y$ is a partial function with domain $\mi{dom}(f) \subseteq X$.}
\item 
Relation
$\preceq \  \subseteq {\cal U}_{E} \times {\cal U}_{E}$ defines a total order on events.\footnote{A total order is a binary relation that is
(1) antisymmetric, i.e.\ $e_1 \preceq e_2$ and $e_2 \preceq e_1$ implies $e_1 = e_2$,
(2) transitive, i.e.\ $e_1 \preceq e_2$ and $e_2 \preceq e_3$ implies $e_1 \preceq e_3$, and
(3) total, i.e., $e_1 \preceq e_2$ or $e_2 \preceq e_1$.}
\end{compactitem}
\end{definition}

Unlike traditional event logs \cite{process-mining-book-2016} we do \emph{not} assume an explicit case notion.
Normally, each event corresponds to precisely one case, i.e., a process instance.
In the \emph{Object-Centric Behavioral Constraint models} (OCBC models) described in Section~\ref{sec:join-two-worlds}
we do not make this restriction and can express \emph{complex interactions between a variety of objects in a single diagram}.
However, to gently introduce the concepts, we first define \emph{constraints over a collection of ordered events}.

\begin{definition}[Event Notations]
Let $E \subseteq {\cal U}_{E}$ be a set of events ordered by $\preceq$ and related
to activities through function $\mi{act}$. For any event $e\in E$:
\begin{compactitem}
  \item $\eincbefore{e}{E} = \{ e' \in E \mid e' \preceq e\}$ are the events \emph{before} and \emph{including} $e$.
  \item $\eincafter{e}{E} = \{ e' \in E \mid e \preceq e'\}$ are the events \emph{after} and \emph{including} $e$.
  \item $\ebefore{e}{E} = \{ e' \in E \mid e' \prec e\}$ are the events \emph{before} $e$.\footnote{$e' \prec e$ if and only if $e' \preceq e$ and $e' \neq e$.}
  \item $\eafter{e}{E} = \{ e' \in E \mid e \prec e'\}$ are the events \emph{after} $e$.
  \item $\partial_a(E) = \{ e' \in E \mid \mi{act}(e') = a\}$ are the events corresponding to activity $a \in {\cal U}_{A}$.
\end{compactitem}
\end{definition}

A process model can be viewed as a set of \emph{constraints}.
In a procedural language like Petri nets, places correspond to constraints:
removing a place may allow for more behavior and adding a place can only restrict behavior.
In this paper, we employ a graphical notation inspired by
\emph{Declare} \cite{declareCSRD09}.
Specifically, we provide a formalization of a subset of \emph{Declare} in terms of
\emph{behavioral cardinality constraints}.
This allows us to reason about behavior and data
in a \emph{unified} manner, since both use cardinality constraints. The following cardinality notion will be used to constrain both data and behavior.

\begin{definition}[Cardinalities]
${\cal U}_\mi{Card} = \{ X \subseteq \Nat \mid  X \neq \emptyset\}$ defines the universe of all possible cardinalities.
Elements of~ ${\cal U}_{\mi{Card}}$ specify non-empty sets of integers.
\end{definition}

Cardinalities are often used in data modeling, e.g., Entity-Relationship (ER) models
and UML Class models may include cardinality constraints. Table~\ref{tab:card} lists a few shorthands typically used in such diagrams. For example, ``$1..^\ast$'' denotes any positive integer.
\begin{SCtable}
\centering
\caption{Some examples of frequently used shorthands for elements of ${\cal U}_{\mi{Card}}$.}\label{tab:card}
\begin{tabular}{|c|c|}
  \hline
 notation & allowed cardinalities \\ \hline
  $1$ & $\{1\}$ \\
  $1..k$ & $\{1,2,\ldots, k\}$ \\
  $^\ast$ & $\{ 0,1,2, \ldots\}$\\
  $1..^\ast$ & $\{1,2, \ldots\}$\\
  \hline
\end{tabular}
\end{SCtable}

In line with literature, we adopt the notation in
Table~\ref{tab:card} for cardinality constraints over data. For
behavioral cardinality constraints, we adopt a different
notation, but very similar in spirit.
Given some \emph{reference event} $e$ we can reason about the events \emph{before} $e$ and the events \emph{after} $e$. We may require that the cardinality of the set of events corresponding to a particular activity before or after the reference event lies within a particular range.
\begin{figure}[t]
\begin{center}
\includegraphics[width=9cm]{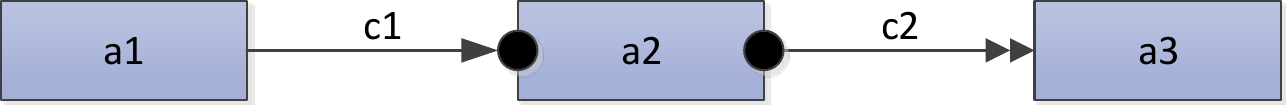}
\caption{Two behavioral cardinality constraints:
constraint $c1$ specifies that all $a2$ events should be preceded by precisely one $a1$ event and constraint $c2$ specifies that all $a2$ events should be succeeded by at least one $a3$ event.}\label{fig-intro-constr}
\end{center}
\end{figure}

Consider for example the two constraints depicted in Figure~\ref{fig-intro-constr}.
Assume a set of events $E \subseteq {\cal U}_{E}$.
The \emph{reference events} for $c1$ are all $a2$ events, i.e.,
all events $E_{\mathit{ref}}^{c1}= \partial_{a2}(E)$. This is indicated by the black dot connecting the $c1$ arrow to activity $a2$.
Now consider a reference event $e_{\mathit{ref}} \in E_{\mathit{ref}}^{c1}$. The single-headed arrow towards the black dot indicates that $e_{\mathit{ref}}$ should be preceded by precisely one $a1$ event.
The $a1$ events are called \emph{target events} (counterpart of $e_{\mathit{ref}}$ when evaluating the constraint).
Formally, constraint $c1$ demands that $|\ebefore{e_{\mathit{ref}}}{\partial_{a1}(E)}| = 1$,
i.e., there has to be precisely one $a1$ event before $e_{\mathit{ref}}$.

The reference events for $c2$ are also all $a2$ events, i.e.,
$E_{\mathit{ref}}^{c2}=\partial_{a2}(E)$. Again, this is visualized by the black dot on the $a2$-side of the constraint.
The double-headed arrow leaving the black dot specifies
that any $e_{\mathit{ref}} \in E_{\mathit{ref}}^{c2}$ should be
followed by at least one $a3$ event. The target events in the context
of $c2$ are all $a3$ events.
Formally: $|\eafter{e_{\mathit{ref}}}{\partial_{a3}(E)}| \geq  1$, i.e.,
there has to be at least one $a3$ event after $e_{\mathit{ref}}$.

The two constraints in Figure~\ref{fig-intro-constr} are just examples.
We allow for any constraint that can be specified in terms of the \emph{cardinality of
preceding and succeeding target events relative
to a collection of reference events}.
Therefore, we define the more general notion of \emph{constraint types}.
\begin{table}[t]
\caption{Examples of constraints types (i.e., elements of ${\cal U}_{\mi{CT}}$), inspired by \emph{Declare}. Note that a constraint is defined with respect of a reference event $e_{\mathit{ref}}$.}\label{tab:exconstrtypes}
\centering
\begin{tabular}{|c|c|}
  \hline
 constraint & formalization \\ \hline
  response & $\{(\mi{before},\mi{after})\in \Nat \times \Nat \mid \mi{after} \geq 1\}$ \\
  unary-response & $\{(\mi{before},\mi{after})\in \Nat \times \Nat \mid \mi{after} = 1\}$ \\
  non-response & $\{(\mi{before},\mi{after})\in \Nat \times \Nat \mid \mi{after} = 0\}$ \\
  precedence & $\{(\mi{before},\mi{after})\in \Nat \times \Nat \mid \mi{before} \geq 1\}$ \\
  unary-precedence & $\{(\mi{before},\mi{after})\in \Nat \times \Nat \mid \mi{before} = 1\}$ \\
  non-precedence & $\{(\mi{before},\mi{after})\in \Nat \times \Nat \mid \mi{before} = 0\}$ \\
  co-existence & $\{(\mi{before},\mi{after})\in \Nat \times \Nat \mid \mi{before}+\mi{after} \geq 1\}$\\
  non-co-existence & $\{(\mi{before},\mi{after})\in \Nat \times \Nat \mid \mi{before}+\mi{after} = 0\}$\\
  \hline
\end{tabular}
\end{table}

\begin{definition}[Constraint Types]
${\cal U}_{\mi{CT}} = \{ X \subseteq \Nat \times \Nat \mid  X \neq \emptyset\}$ defines the universe of all possible constraint types. Any element of ${\cal U}_{\mi{CT}}$ specifies a non-empty set of pairs of integers: the first integer defines the number of target events before the reference event and the second integer defines the number of target events after the reference event.
\end{definition}
\begin{figure}[t]
\begin{center}
\includegraphics[width=12cm]{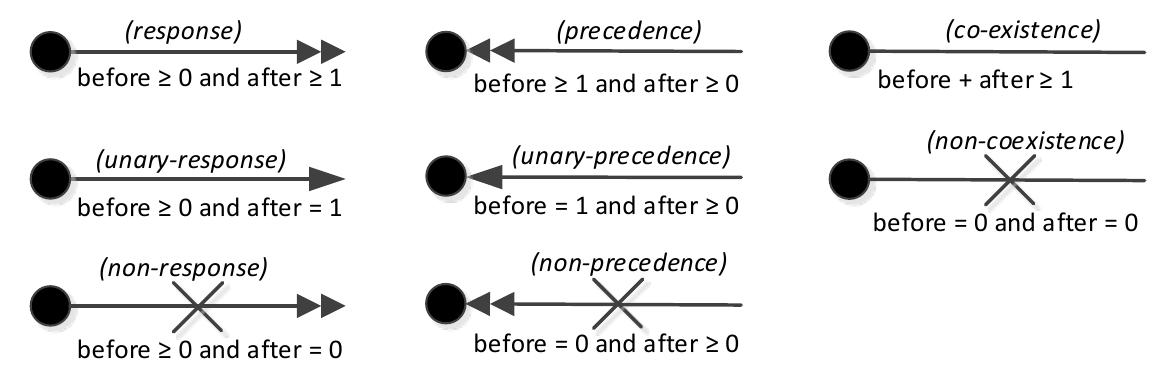}
\caption{Graphical notation for the example constraint types defined in Table~\ref{tab:exconstrtypes} (example elements of ${\cal U}_{\mi{CT}}$).
The dot on the left-hand side of each constraint refers to the \emph{reference events}.
\emph{Target events} are on the other side that has no dot.
The notation is inspired by \emph{Declare}, but formalized in terms of
cardinality constraints rather than LTL.}\label{fig-arrow-notation}
\end{center}
\end{figure}

\begin{figure}[t]
\begin{center}
\includegraphics[width=12cm]{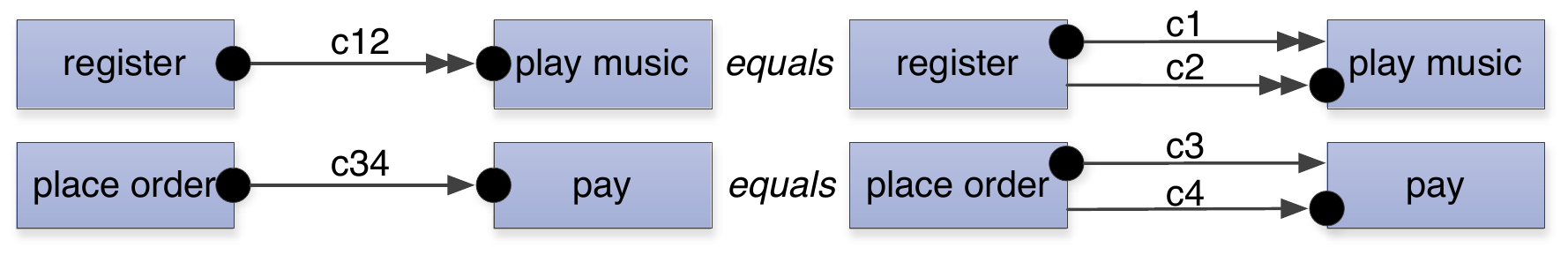}
\caption{An arrow with two reference events ($\bullet$) can be used as a shorthand. Constraint $c12$ ($c34$) corresponds to the conjunction of constraints $c1$ and $c2$ (resp.\ $c3$ and $c4$).}\label{fig-arrow-shorthands}
\end{center}
\end{figure}

Table~\ref{tab:exconstrtypes} shows eight examples of constraint types.
Constraint $c1$ is a unary-precedence constraint and constraint $c2$ is a response constraint. The graphical representations of the eight example constraint types are shown in Figure~\ref{fig-arrow-notation}.
As a shorthand, one arrow may combine two constraints as shown in Figure~\ref{fig-arrow-shorthands}.
For example, constraint $c34$ states that after placing an order there is precisely one payment and before a payment there is precisely one order placement.

A \emph{Behavioral Constraint} (BC) model is a collection of activities and constraints (cf.\ Figure~\ref{fig-intro-constr}).\footnote{Note that BC models depend on the classical instance notion, i.e., the model describes the lifecycle of single instance. OCBC models do not assume a single instance notion. However, we introduce behavioral constraint models to gently introduce the concepts.}

\begin{definition}[Behavioral Constraint Model]\label{def:bcm}
A \emph{behavioral constraint model} is a tuple
$\mi{BCM} = (A,C,\pi_{\mathit{ref}},\pi_{\mathit{tar}}, \allowbreak
\mi{type})$, where
\begin{compactitem}
  \item $A\subseteq {\cal U}_{A}$ is the set of activities (denoted by rectangles),
  \item $C$ is the set of constraints ($A \cap C = \emptyset$, denoted by various types of edges),
  \item $\pi_{\mathit{ref}} \in C \rightarrow A$ defines the reference activity of a constraint (denoted by a black dot connecting constraint and activity),
  \item $\pi_{\mathit{tar}} \in C \rightarrow A$ defines the target activity of a constraint (other side of edge), and
  \item $\mi{type} \in C \rightarrow {\cal U}_{\mi{CT}}$ specifies the type of each constraint (denoted by the type of edge).
\end{compactitem}
\end{definition}

Figure~\ref{fig-intro-constr} defines the BC model
$\mi{BCM} = (A,C,\pi_{\mathit{ref}},\pi_{\mathit{tar}}, \allowbreak \mi{type})$ with
$A=\{a1,a2,a3\}$,
$C=\{c1,c2\}$,
$\pi_{\mathit{ref}}(c1) =  a2$, $\pi_{\mathit{ref}}(c2) = a2$,
$\pi_{\mathit{tar}}(c1) = a1$, $\pi_{\mathit{tar}}(c2) = a3$,
$\mi{type}(c1) = \{(\mi{before},\mi{after})\in \Nat \times \Nat \mid \mi{before} = 1\}$ (unary-precedence), and
$\mi{type}(c2) = \{(\mi{before},\mi{after})\in \Nat \times \Nat \mid \mi{after} \geq 1\}$ (response).
Given a set $E$ of events, we can check whether constraints are satisfied (or not), thus providing a natural link to conformance checking.

\begin{definition}[Constraint Satisfaction]\label{def:bcmsc}
Let $\mi{BCM} = (A,C,\pi_{\mathit{ref}},\pi_{\mathit{tar}},\mi{type})$ be a BC model,
and $E \subseteq {\cal U}_{E}$ a set of events.
\begin{compactitem}
\item Event set $E$ \emph{satisfies} constraint $c$ if and only if
  \[(|\ebefore{e_{\mathit{ref}}}{\partial_{\pi_{\mathit{tar}}(c)}(E)}|,|\eafter{e_{\mathit{ref}}}{\partial_{\pi_{\mathit{tar}}(c)}(E)}|) \in  \mi{type}(c)
\textrm{ for all } e_{\mathit{ref}} \in \partial_{\pi_{\mathit{ref}}(c)}(E)\]
\item Event set $E$ satisfies $\mi{BCM}$ if and only if $E$ satisfies each constraint $c \in C$.
\end{compactitem}
\end{definition}

The reference activity of a constraint defines the corresponding set of reference events $E_{\mathit{ref}}^c$. For each reference event it is checked whether the cardinality constraint is satisfied. $\ebefore{e_{\mathit{ref}}}{\partial_{\pi_{\mathit{tar}}(c)}(E)}$ are all target events before the reference event $e_{\mathit{ref}}$. $\eafter{e_{\mathit{ref}}}{\partial_{\pi_{\mathit{tar}}(c)}(E)}$ are all target events after the reference event $e_{\mathit{ref}}$.
Consider, e.g., $c1$ in Figure~\ref{fig-intro-constr}.
All $a2$ events are reference events and all $a1$ events are target events.
For this example $\ebefore{e_{\mathit{ref}}}{\partial_{\pi_{\mathit{tar}}(c1)}(E)}$ is the set of $a1$ events before
the selected $a2$ event ($e_{\mathit{ref}}$). The cardinality of this set should be precisely 1.

In traditional process modeling notations a constraint is defined for one process instance (case) in isolation.
This means that the set $E$ in Definition~\ref{def:bcmsc} refers to all events corresponding to the same case.
As discussed before, the case notion is often too rigid.
There may be multiple case-notions at the same time, causing one-to-many or many-to-many relations
that cannot be handled using traditional monolithic process models.
Moreover, we need to relate events to (data) objects. All these issues are discussed next.

\section{Modeling Data Cardinality Constraints}\label{sec:dataconstr}

\begin{figure}[tbh!]
\begin{center}
\includegraphics[width=12cm]{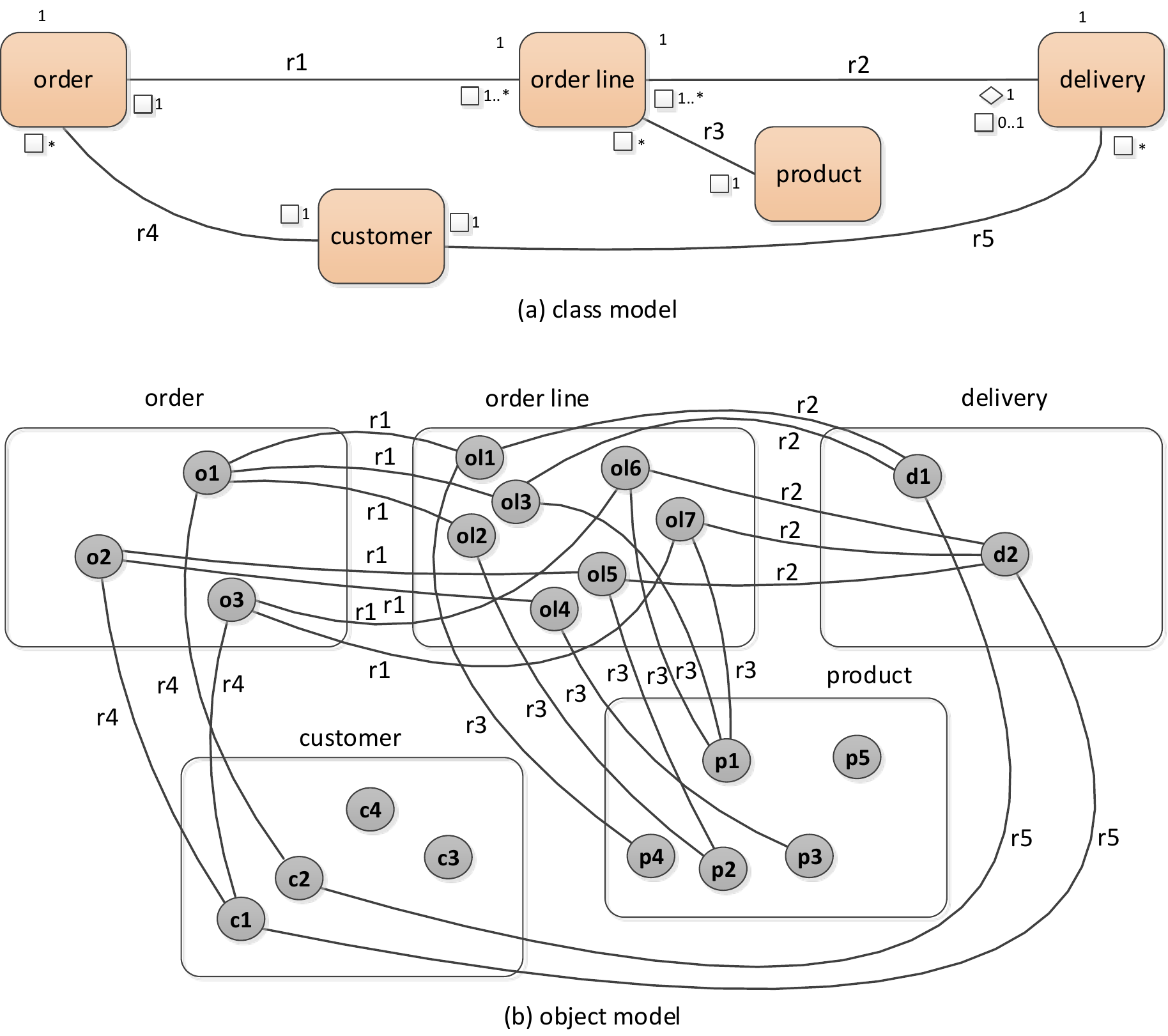}
\caption{Example of a class model and corresponding object model.}\label{fig-class-model}
\end{center}
\end{figure}

Next to behavior as captured through events, there are also objects that are grouped in classes.
Objects may be related and cardinality constraints help to structure dependencies.
Entity-Relationship (ER) models \cite{ER_Chen76}, UML class models \cite{uml_omg_2_5}, and Object-Role Models (ORM) \cite{Halpin-ORM-book} are examples
of notations used for object modeling, often referred to as data modeling.
In this paper, we use the simple notation shown in Figure~\ref{fig-class-model}(a) to specify \emph{class models}.
The notation can be viewed as a subset of such mainstream notations.
The only particular feature is that cardinality constraints
can be tagged as ``always'' ($\alwys$) or ``eventually'' ($\eventy$).
For example, for every order order line there is always at most one delivery ($\alwys~0..1$) and eventually (i.e., from some point in time onwards) there
should be a corresponding delivery($\eventy~1$).

\begin{definition}[Class Model]\label{def:classmod}
A \emph{class model} is a tuple $\mi{ClaM} = (\mi{OC},\mi{RT},\pi_1,\allowbreak \pi_2,\allowbreak \cardsrcalways,\allowbreak \cardsrceven,\allowbreak \cardtaralways,\allowbreak \cardtareven)$,where
\begin{compactitem}
  \item $\mi{OC}$ is a set of object classes,
  \item $\mi{RT}$ is a set of relationship types ($\mi{OC} \cap \mi{RT} = \emptyset$),
  \item $\pi_1 \in \mi{RT}\rightarrow \mi{OC}$ gives the source of a relationship,
  \item $\pi_2 \in \mi{RT} \rightarrow \mi{OC}$ gives the target of a relationship,
\item $\cardsrcalways \in RT \rightarrow \mathcal{U}_{Card}$ gives the cardinality of the source of a relationship (the constraint should hold at any point in time as indicated by $\alwys$),
\item $\cardsrceven \in RT \rightarrow \mathcal{U}_{Card}$ gives the cardinality of the source of a relationship (the constraint should hold from some point onwards as indicated by $\eventy$),
\item $\cardtaralways \in RT \rightarrow \mathcal{U}_{Card}$ gives the cardinality of the target of a relationship (the constraint should hold at any point in time as indicated by $\alwys$), and
\item $\cardtareven \in RT \rightarrow \mathcal{U}_{Card}$ gives the cardinality of the target of a relationship (the constraint should hold from some point onwards as indicated by $\eventy$).
      \end{compactitem}
\end{definition}

The class model $\mi{ClaM} = (\mi{OC},\mi{RT},\pi_1,\allowbreak \pi_2,\allowbreak \cardsrcalways,\allowbreak \cardsrceven,\allowbreak \cardtaralways,\allowbreak \cardtareven)$ depicted in
Figure~\ref{fig-class-model}(a) has five object classes
$\mi{OC} = \{ \mi{order}, \mi{order~line}, \mi{delivery}, \mi{customer}, \mi{product} \}$ and five
relationship types $\mi{RT} = \{ \mi{r1}, \mi{r2}, \mi{r3}, \mi{r4}, \mi{r5} \}$.
Relationship type $\mi{r1}$ is connecting classes $\mi{order}$ and $\mi{order~line}$:
$\pi_1(r1) = \mi{order}$ and $\pi_2(r1) = \mi{order~line}$.
For the other relationships types, we have:
$\pi_1(r2) = \mi{order~line}$, $\pi_2(r2) = \mi{delivery}$,
$\pi_1(r3) = \mi{order~line}$, and $\pi_2(r3) = \mi{product}$, etc.

The notation of Table~\ref{tab:card} is extended with $\alwys$ (``always'') or $\eventy$ (``eventually'') to specify the cardinalities in Figure~\ref{fig-class-model}(a).
$\cardsrcalways(r1)= \{ 1\}$, i.e., for each object in class $\mi{order~line}$ there is always precisely one corresponding object in $\mi{order}$.
This is indicated by the ``$\alwys~1$'' annotation on the source side (i.e., the $\mi{order}$ side of $r1$) in Figure~\ref{fig-class-model}(a).
$\cardtaralways(r1)= \{ 1,2,3, \ldots \}$, i.e., for each object in class $\mi{order}$ there is always at least one corresponding object in $\mi{order~line}$.
This is indicated by the ``$\alwys~1..^\ast$'' annotation on the target side (i.e., the $\mi{order~line}$ side) of $r1$.
Not shown are $\cardsrceven(r1)= \{ 1\}$ (``$\eventy~1$'') and $\cardtareven(r1)= \{ 1,2,3, \ldots \}$ (``$\eventy~1..^\ast$'') as these are implied by the ``always'' constraints.
One the target side of $r2$ in Figure~\ref{fig-class-model}(a) there are two
cardinality constraints: $\cardtaralways(r2)= \{ 0,1\}$ and $\cardtareven(r2)= \{ 1\}$.
This models that eventually each order line needs to have a corresponding delivery (``$\eventy~1$''). However, the corresponding delivery may be created later (``$\alwys~0..1$'').
We only show the ``eventually'' ($\eventy$) cardinality constraints that are more restrictive than the ``always'' ($\alwys$) cardinalities in the class model. Obviously, $\cardsrceven(r) \subseteq \cardsrcalways(r)$
and $\cardtareven(r) \subseteq \cardtaralways(r)$ for any $r \in RT$
since constraints that always hold also hold eventually.

Objects can also have attributes and therefore in principle the class model should list the names and types of these attributes.
We abstract from object/class attributes in this paper, as well as
from the notions of hierarchies and subtyping,
but they could be added in a straightforward manner.

A class diagram defines a ``space'' of possible \emph{object models}, i.e., concrete collections of objects and relations instantiating the class model.

\begin{definition}[Object Model]\label{def:objmode}
${\cal U}_{O}$ is the universe of object identifiers.
An object model for class model $\mi{ClaM} = (\mi{OC},\mi{RT},\pi_1,\allowbreak \pi_2,\allowbreak \cardsrcalways,\allowbreak \cardsrceven,\allowbreak \cardtaralways,\allowbreak \cardtareven)$ is
a tuple $\mi{OM} = (\mi{Obj},\mi{Rel},\mi{class})$, where:
\begin{compactitem}
  \item $\mi{Obj}  \subseteq {\cal U}_{O}$ is the set of objects,
  \item $\mi{Rel} \subseteq  \mi{RT} \times \mi{Obj}  \times \mi{Obj}$ is the set of relations,
  \item $\mi{class} \in \mi{Obj} \rightarrow \mi{OC}$ maps objects onto classes.
      \end{compactitem}
${\cal U}_{\mi{OM}}$ is the universe of object models.
\end{definition}

Figure~\ref{fig-class-model}(b) shows an object model $\mi{OM} = (\mi{Obj},\mi{Rel},\mi{class})$.
The  objects are depicted as grey dots:
$\mi{Obj} = \{\mi{o1},\mi{o2}, \mi{o3},\allowbreak \mi{ol1},\allowbreak \mi{ol2},\allowbreak  \ldots,\allowbreak   \mi{ol7},\allowbreak  \mi{d1},\allowbreak  \mi{d2},\allowbreak  \mi{c1}, \ldots,  \mi{c4}, \mi{p1}, \ldots,  \mi{p5} \}$.
There are three objects belonging to object class $\mi{oc1}$, i.e.,
$\mi{class}(\mi{o1}) = \mi{class}(\mi{o2}) = \mi{class}(\mi{o3}) = \mi{order}$.
There are seven relations corresponding to relationship $\mi{r1}$, e.g.,
$(\mi{r1},\mi{o1},\mi{ol1}) \in \mi{Rel}$ and $(\mi{r1},\mi{o2},\mi{ol5}) \in \mi{Rel}$.

Note that objects and events are represented by unique identifiers.
This allows us to refer to a specific object or event.
Even two events or objects with the same properties are still distinguishable by their identity.

The cardinalities specified in the class model should be respected by the object model. For example, for each object in class $\mi{order~line}$ there is precisely one corresponding object in $\mi{order}$ according to $\mi{r1}$. A \emph{valid} object model complies with the ``always'' ($\alwys$) cardinalities in the class model. A valid model is also \emph{fulfilled} is also the possibly stronger ``eventually'' ($\eventy$) cardinality constraints are satisfied.

\begin{definition}[Valid Object Model]\label{def:valopmod}
Let $\mi{ClaM} = (\mi{OC},\mi{RT},\pi_1,\allowbreak \pi_2,\allowbreak \cardsrcalways,\allowbreak \cardsrceven,\allowbreak \cardtaralways,\allowbreak \cardtareven)$ be a class model and
$\mi{OM} = (\mi{Obj},\mi{Rel},\mi{class}) \in {\cal U}_{\mi{OM}}$ be an object
model. $\mi{OM}$ is valid for $\mi{ClaM}$ if and only if
\begin{compactitem}
  \item for any $(r,o_1,o_2) \in \mi{Rel}$: $\mi{class}(o_1) = \pi_1(r)$ and $\mi{class}(o_2) = \pi_2(r)$,
  \item for any $r \in \mi{RT}$ and $o_2
    \in \partial_{\pi_2(r)}(\mi{Obj})$, we have that \footnote{$\partial_{\mi{oc}}(\mi{Obj}) = \{ o
          \in \mi{Obj} \mid \mi{class}(o) = \mi{oc}\}$ denotes the whole
          set of objects in class $\mi{oc}$.}
        \[|\{ o_1 \in \mi{Obj} \mid (r,o_1,o_2) \in \mi{Rel}  \}| \in
        \cardsrcalways(r), \textrm{ and}
\]
  \item for any $r \in \mi{RT}$ and $o_1
    \in \partial_{\pi_1(r)}(\mi{Obj})$, we have that
        \[|\{ o_2 \in \mi{Obj} \mid (r,o_1,o_2) \in \mi{Rel} \}| \in \cardtaralways(r)\]
      \end{compactitem}
A valid objected model is also fulfilled if the stronger cardinality constraints hold (these are supposed to hold eventually):
\begin{compactitem}
  \item for any $r \in \mi{RT}$ and $o_2
    \in \partial_{\pi_2(r)}(\mi{Obj})$, we have that
        \[|\{ o_1 \in \mi{Obj} \mid (r,o_1,o_2) \in \mi{Rel}  \}| \in
        \cardsrceven(r), \textrm{ and}
\]
  \item for any $r \in \mi{RT}$ and $o_1
    \in \partial_{\pi_1(r)}(\mi{Obj})$, we have that
        \[|\{ o_2 \in \mi{Obj} \mid (r,o_1,o_2) \in \mi{Rel} \}| \in \cardtareven(r)\]
      \end{compactitem}
\end{definition}

The object model in Figure~\ref{fig-class-model}(b) is indeed valid.
If we would remove relation $(\mi{r1},\mi{o1},\mi{ol1})$,
the model would no longer be valid
(because an order line should always have a corresponding order).
Adding a relation $(\mi{r1},\mi{o2},\mi{ol1})$ would also destroy validity.
Both changes would violate the ``$\alwys~1$'' constraint on the
source side of $r1$.
The object model in Figure~\ref{fig-class-model}(b) is
not fulfilled because the ``$\eventy~1$'' constraint on the
target side of $r2$ does not hold.
Order lines $\mi{ol2}$ and $\mi{ol4}$ do not (yet) have a corresponding delivery.
Adding deliveries for these order lines and adding the corresponding relations
would make the model fulfilled.

Definition~\ref{def:valopmod} only formalizes simple cardinality constraints involving a binary relation and abstracting from attribute values.
In principle more sophisticated constraints could be considered: the object model $\mi{OM}$ is simply checked against a class model $\mi{ClaM}$.
For example, the Object Constraint Language (OCL) \cite{omg_ocl_2_4} could be used to define more refined constraints.

\section{Object-Centric Behavioral Constraints}\label{sec:join-two-worlds}

In Section~\ref{sec:behavconstr}, we focused on control-flow modeling and formalized behavioral constraints \emph{without} considering the structure of objects.
In Section~\ref{sec:dataconstr}, we focused on structuring objects and formalized cardinality constraints on object models (i.e., classical data modeling).
In this section, we combine \emph{both perspectives} to fully address the challenges
described in the introduction.

\subsection{Object-Centric Event Logs}

First, we formalize the notion of an event log building on the event notion introduced in Definition~\ref{def:events}.
An event log is a collection of events that belong together, i.e., they belong to some ``process'' where many types of objects/instances may interact.
Next to scoping the log, we also relate events to objects.
Note that the same event may refer to multiple objects and one object may be referred to by multiple events.

\begin{definition}[Event Log]\label{def:eventlog}
An \emph{event log} is a tuple $L = (E,\allowbreak \mi{act},\allowbreak \mi{attr}, \mi{EO}, \mi{om},
\preceq)$, where
\begin{compactitem}
  \item $E \subseteq {\cal U}_{E}$ is a set of events,
  \item $\mi{act} \in E \rightarrow {\cal U}_{A}$ maps events onto activities,
  \item $\mi{attr} \in E \rightarrow ( {\cal U}_{\mi{Attr}} \not\rightarrow {\cal U}_{\mi{Val}})$ maps events onto a partial function assigning values to some attributes,
  \item $\mi{EO} \subseteq E \times {\cal U}_{O}$ relates events to sets of object references,
  \item $\mi{om} \in E \rightarrow {\cal U}_{\mi{OM}}$ maps each event to the object model directly after the event took place, and
  \item $\preceq \  \subseteq E \times E$ defines a total order on events.
\end{compactitem}
\end{definition}

In the context of an event $L$, each event $e$ is associated with object model $\mi{OM}_{e} = (\mi{Obj}_{e},\mi{Rel}_{e},\mi{class}_{e})  = \mi{om}(e)$.
In the remainder, we refer directly to $\mi{Obj}_{e}$, $\mi{Rel}_{e}$, $\mi{class}_{e}$ for $e \in E$ if the context is clear.
\begin{figure}
\begin{center}
\includegraphics[width=8cm]{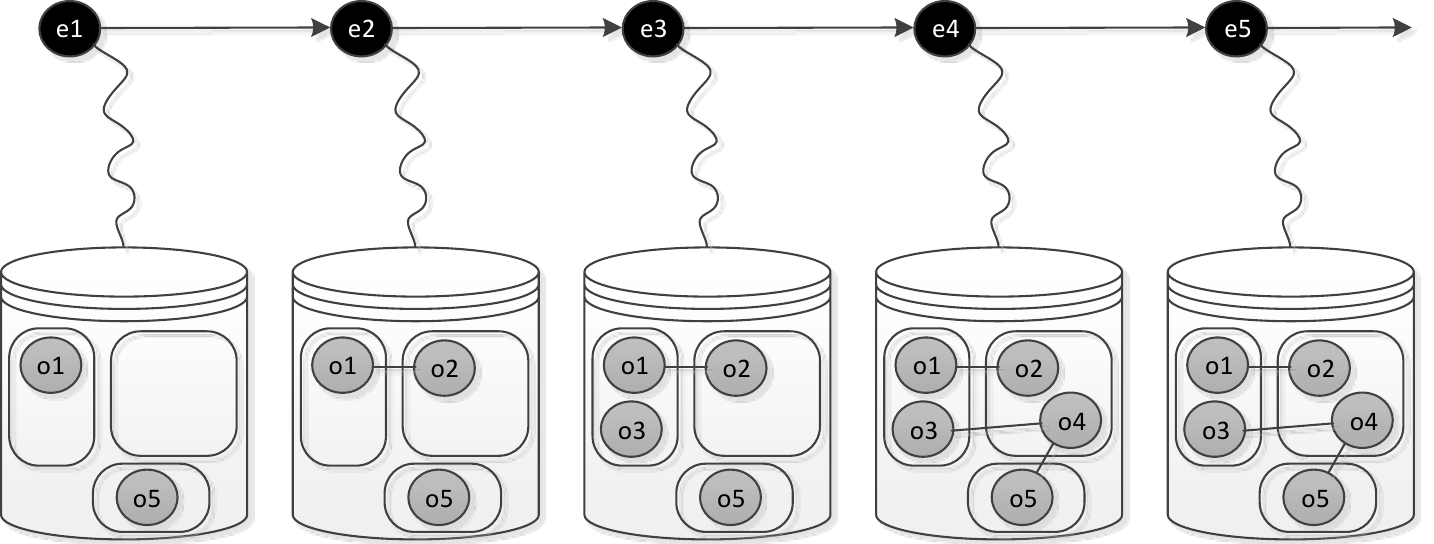}
\caption{Each event $e$ refers to the object model right after $e$ occurred: $\mi{OM}_{e} = (\mi{Obj}_{e},\mi{Rel}_{e},\mi{class}_{e})  = \mi{om}(e)$.}\label{fig-om-evo}
\end{center}
\end{figure}

Figure~\ref{fig-om-evo} illustrates the evolution of the object model.
After the occurrence of some event $e$ objects may have been added (we assume monotonicity), and
relationships may have been added or removed.
Event $e$ may refer to objects through relation $\mi{EO}$ and these objects need to exist,
i.e., for all $(e,o) \in \mi{EO}$: $o \in \mi{Obj}_{e}$.
We assume that objects cannot be removed at a later stage to avoid referencing non-existent objects.
Objects can be marked as deleted but cannot be removed (e.g, by using an attribute or relation).

The event log provides a \emph{snapshot of the object model after each event}.
This triggers the question: Can the object model be changed in-between two subsequent events?
If no such changes are possible,
then the object model before an event is the same as the object model after the previous event.
If we would like to allow for updates in-between events, then these could be recorded in the log.
Events referring to some artificial activity \emph{update} could be added to signal the updated object model.
We could also explicitly add a snapshot of the object model just before each event.
In the remainder, we only consider the
snapshot $\mi{OM}_{e}$ after each event $e\in E$.

Note that Definition~\ref{def:eventlog} calls for event logs different
from the standard \emph{XES format}. XES (\url{www.xes-standard.org}), which is supported by the majority of process mining tools, assumes a case notion (i.e.,
each event refers to a process instance) and does not keep track of object models.


\subsection{OCBC Models}

Next, we define \emph{Object-Centric Behavioral Constraint} (OCBC) models.
Through a combination of control-flow modeling and data/object modeling, we relate behavior and structure.
The BC models from Section~\ref{sec:behavconstr}
are connected to the class models of Section~\ref{sec:dataconstr}
to provide the integration needed.
\begin{figure}[tbh!]
\begin{center}
\includegraphics[width=10cm]{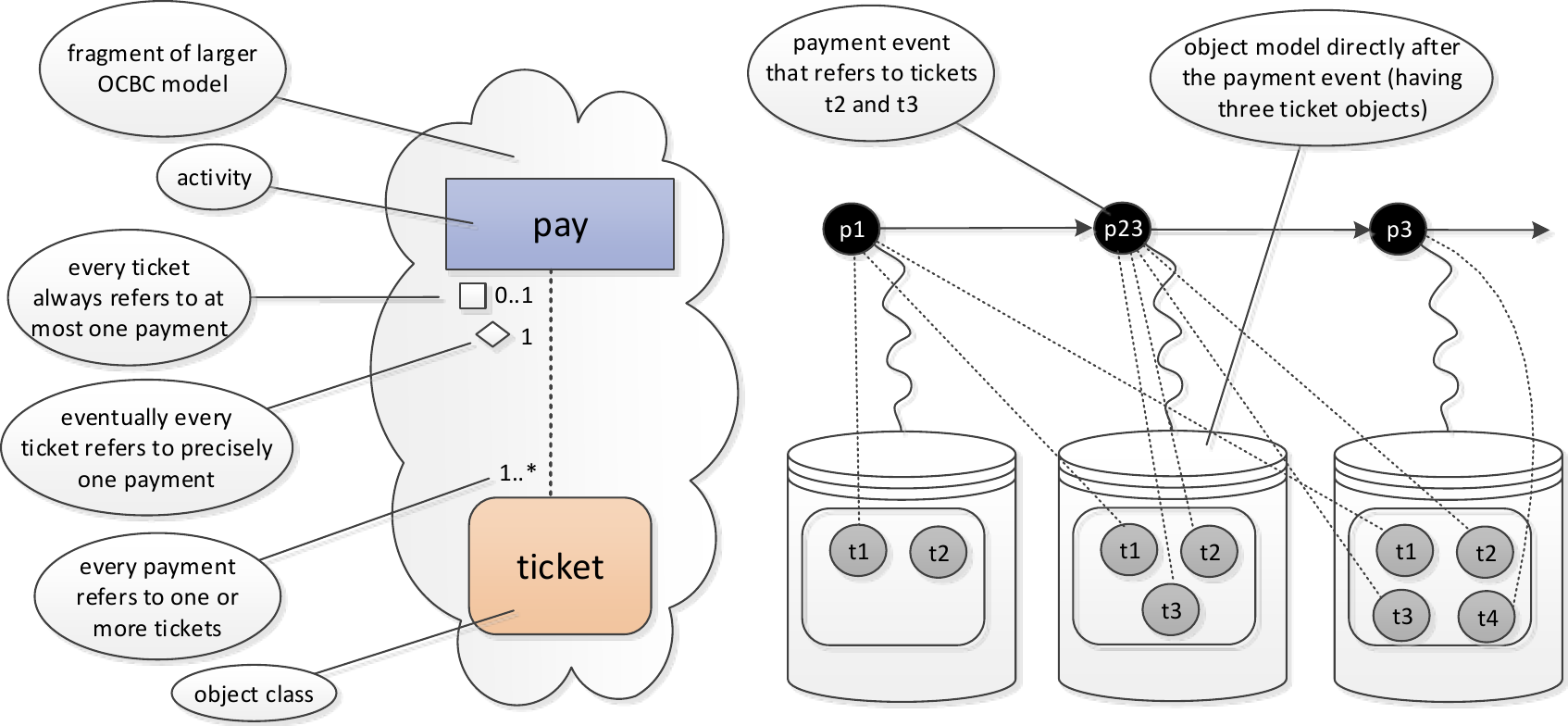}
\caption{Illustrating cardinality constraints $\acardalways$, $\acardeven$, and $\ocard$.}\label{fig-constraint-eval}
\end{center}
\end{figure}

A key ingredient is that events and objects are related as illustrated in Figure~\ref{fig-constraint-eval}.
Payment activity $\mi{p1}$ refers to ticket $\mi{t1}$,
activity $\mi{p23}$ refers to tickets $\mi{t2}$ and $\mi{t3}$,
and activity $\mi{p3}$ refers to ticket $\mi{t4}$.
Figure~\ref{fig-constraint-eval} shows three example constraints:
``$\alwys~0..1$'' (every ticket always refers to at most one payment),
``$\eventy~1$'' (eventually every ticket refers to precisely one payment),
and ``$1..^\ast$'' (every payment refers to one or more tickets).

\begin{definition}[Object-Centric Behavioral Constraint Model]\label{def:ocbcmodel} An
  \emph{object-centric behavioral constraint model} is a tuple
$\mi{OCBCM}$ $=$
$(\mi{BCM},\mi{ClaM},\mi{AOC},\acardalways, \acardeven, \ocard, \mi{crel})$, where
\begin{compactitem}
  \item $\mi{BCM} = (A,C,\pi_{\mathit{ref}},\pi_{\mathit{tar}},\mi{type})$ is a BC model (Definition~\ref{def:bcm}),
  \item $\mi{ClaM} = (\mi{OC},\mi{RT},\pi_1,\allowbreak \pi_2,\allowbreak \cardsrcalways,\allowbreak \cardsrceven,\allowbreak \cardtaralways,\allowbreak \cardtareven)$ is a class model (Definition~\ref{def:classmod}),
  \item $A$, $C$, $OC$ and $RT$ are pairwise disjoint (no name clashes),
  \item $\mi{AOC} \subseteq A \times \mi{OC}$ is a set of relations between activities and object classes,
  \item $\acardalways \in \mi{AOC} \rightarrow {\cal U}_{\mi{Card}}$ gives the cardinality of the source of a relation linking an activity and an object class (activity side,
   the constraint should hold at any point in time as indicated by $\alwys$),
  \item $\acardeven \in \mi{AOC} \rightarrow {\cal U}_{\mi{Card}}$ gives the cardinality of the source of a relation linking an activity and an object class (activity side, the constraint should hold from some point onwards as indicated by $\eventy$),
  \item $\ocard \in \mi{AOC} \rightarrow {\cal U}_{\mi{Card}}$ gives the cardinality of the target of a relation linking an activity and an object class (object-class side), and
  \item $\mi{crel} \in C \rightarrow \mi{OC} \cup \mi{RT}$ is the constraint relation satisfying the following conditions for each $c\in C$:
      \begin{compactitem}
      \item $\{(\pi_{\mathit{ref}}(c),\mi{oc}),(\pi_{\mathit{tar}}(c),\mi{oc})\} \subseteq \mi{AOC}$ if $\mi{crel}(c) = \mi{oc} \in \mi{OC}$, and
      \item $\{(\pi_{\mathit{ref}}(c),\pi_1(r)),(\pi_{\mathit{tar}}(c),\pi_2(r))\} \subseteq \mi{AOC}$ or $\{(\pi_{\mathit{ref}}(c),\pi_2(r)),\allowbreak (\pi_{\mathit{tar}}(c),\allowbreak \pi_1(r))\}\allowbreak  \subseteq \mi{AOC}$ if $\mi{crel}(c) = r \in \mi{RT}$.
     \end{compactitem}

\end{compactitem}
\end{definition}
\begin{figure}[tbh!]
\begin{center}
\includegraphics[width=11cm]{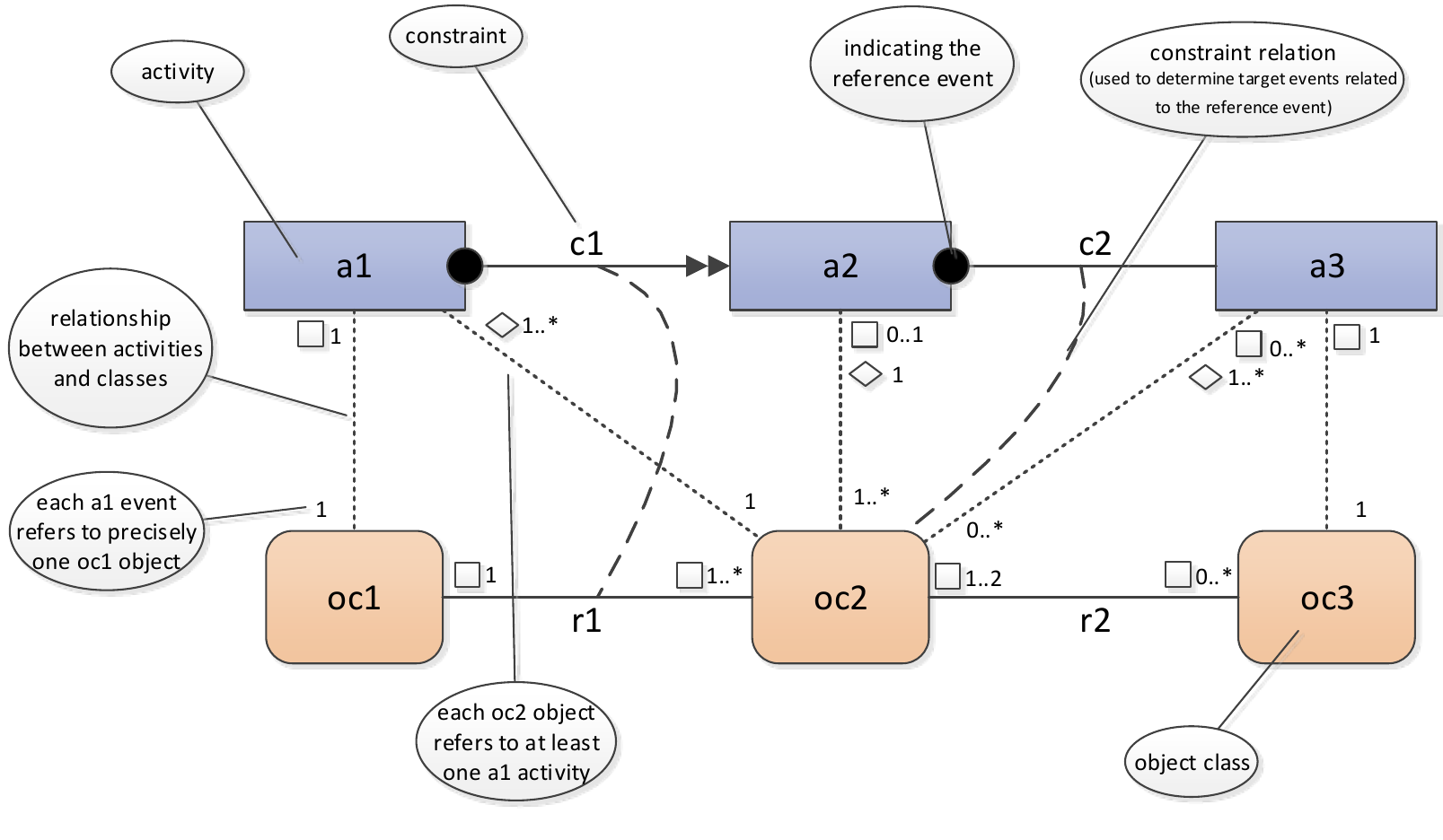}
\caption{An example model illustrating the main ingredients of an OCBC model.}\label{fig-example1}
\end{center}
\end{figure}

An \emph{Object-Centric Behavioral Constraint model} (OCBC model) includes a
behavioral constraint model (to model behavior)
and a class model (to model objects/data).
These are related through relation $\mi{AOC}$ and functions $\acardalways$, $\acardeven$, $\ocard$, and $\mi{crel}$. We use Figure~\ref{fig-example1} to clarify these concepts.

$\mi{AOC}$ relates activities and object classes.
In Figure~\ref{fig-example1}, $\mi{AOC} = \{ (\mi{a1},\mi{oc1}), \allowbreak (\mi{a1},\mi{oc2}), (\mi{a2},\mi{oc2}),(\mi{a3},\mi{oc2}),(\mi{a3},\mi{oc3}) \}$.
For example, $\mi{a1}$ may potentially refer to $\mi{oc1}$ and $\mi{oc2}$ objects,
but not to $\mi{oc3}$ objects because $(\mi{a1},\mi{oc3})\not\in \mi{AOC}$.
Recall that in an event log $L$ there is a many-to-many relationship between events and objects ($\mi{EO} \subseteq E \times {\cal U}_{O}$) constrained by $\mi{AOC}$.

Functions $\acardalways$, $\acardeven$, and $\ocard$ define possible cardinalities,
similar to cardinality constraints in a class model.
Functions $\acardalways$ and $\acardeven$ define how many events there need to be for each object.
Since the object model is evolving, there are two types of constraints: constraints that should hold \emph{at any point in time} from the moment the object exists ($\acardalways$)
and constraints that should \emph{eventually} hold $\acardeven$.
Function $\ocard$ defines how many objects there need to be for each event when the event occurs (specified by $\mi{EO}$).

As indicated by the ``$\alwys~1$'' annotation on the $\mi{a1}$-side of the line
connecting activity $\mi{a1}$ and object class $\mi{oc1}$,
there is precisely one $\mi{a1}$ event for each $\mi{oc1}$ object (from the moment it exists):
$\acardalways(\mi{a1},\mi{oc1}) = \{1\}$.
As indicated by the ``$\eventy~1..^\ast$'' on the $\mi{a1}$-side of the line
connecting activity $\mi{a1}$ and object class $\mi{oc2}$, we have
that $\acardeven(\mi{a1},\mi{oc2}) = \{1,2, \ldots\}$.
This means that \emph{eventually} each $\mi{oc2}$ object refers to at least one $\mi{a1}$ activity.
Note that an $\mi{oc2}$ object does not need to have a corresponding $\mi{a1}$ event when it is created.
However, adding a new $\mi{oc2}$ object implies the occurrence of at least one corresponding $\mi{a1}$ event to satisfy the cardinality constraint ``$\eventy~1..^\ast$'',
i.e., an \emph{obligation} is created.
If the annotation ``$\alwys~1..^\ast$'' would have been used (instead of ``$\eventy~1..^\ast$''), then the creation of any $\mi{oc2}$ object needs to coincide with a corresponding $\mi{a1}$ event,
because the cardinality constraints should always hold ($\alwys$) and not just eventually ($\eventy$).

As indicated by the ``$1$'' annotation on the $\mi{oc2}$-side of the
line connecting activity $\mi{a1}$ and object class $\mi{oc2}$, we
then have that $\ocard(\mi{a1},\mi{oc2}) = \{1\}$. This means that each $\mi{a1}$ activity refers to precisely one $\mi{oc2}$ object.

Let's now consider relation $(\mi{a2},\mi{oc2}) \in \mi{AOC}$.
There should be at most one $\mi{a2}$ event for each $\mi{oc2}$ object from the moment it exists: $\acardalways(\mi{a2},\mi{oc2}) = \{0,1\}$.
Eventually there should be precisely one $\mi{a2}$ event for each $\mi{oc2}$ object: $\acardeven(\mi{a2},\mi{oc2}) = \{1\}$.
$\ocard(\mi{a2},\mi{oc2}) = \{1,2, \ldots\}$ indicates that each $\mi{a2}$ event refers to at least one $\mi{oc2}$ object.

Annotations of the type ``$\eventy~0..^\ast$'' and ``$\alwys~0..^\ast$'' are omitted from the diagram because these impose no constraints.
Also implied constraints can be left out, e.g., ``$\alwys~1..^\ast$'' implies ``$\eventy~1..^\ast$''.
\begin{figure}[t]
\begin{center}
\includegraphics[width=9cm]{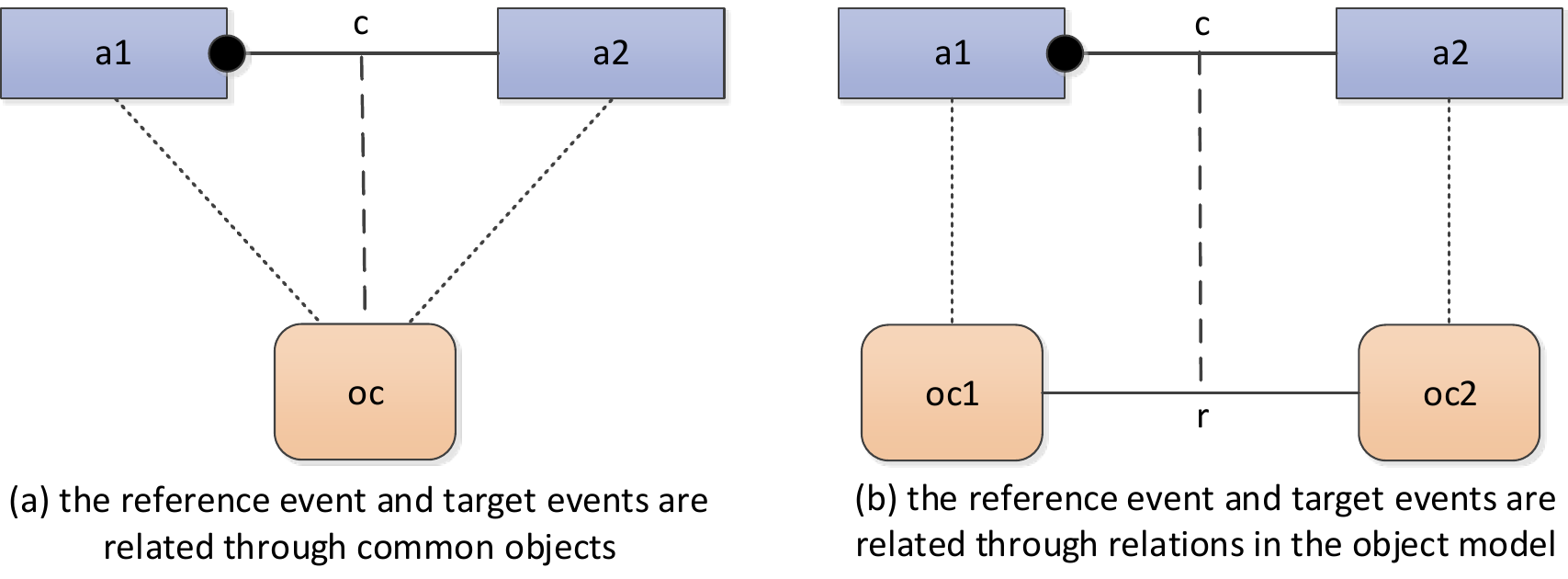}
\caption{Two types of constraint relations:
(a) $\mi{crel}(c) = \mi{oc} \in \mi{OC}$, i.e., the target events are related to the reference event through shared objects of the class $\mi{oc}$,
(b) $\mi{crel}(c) = r \in \mi{RT}$, i.e., the target events are related to the reference event through relations of type $\mi{r}$ (in any direction).}\label{fig-two-type-ofconstraints}
\end{center}
\end{figure}

\begin{figure}[t]
\begin{center}
\includegraphics[width=12cm]{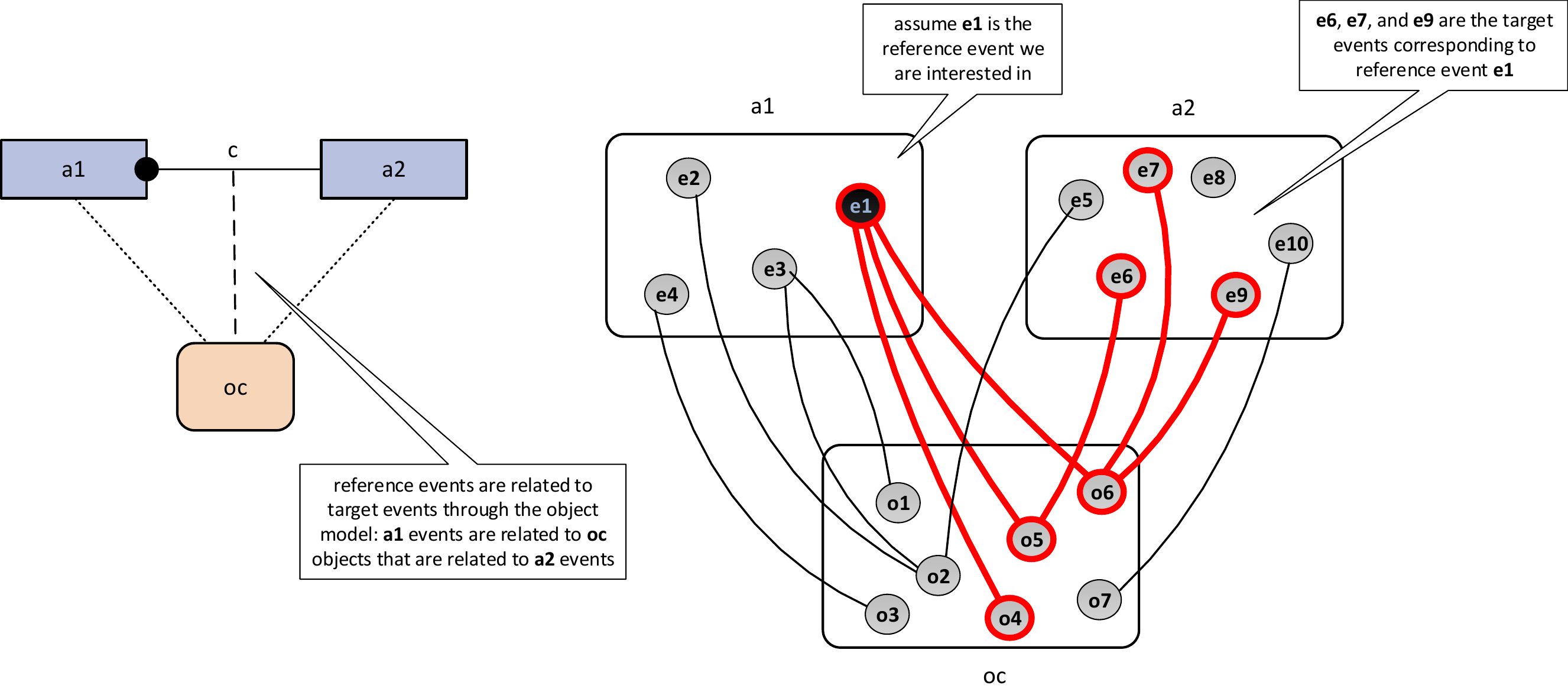}
\caption{Given a reference event for a constraint with
$\mi{crel}(c) = \mi{oc} \in \mi{OC}$ we navigate to the target events through shared object references.}\label{fig-constraint-type1}
\end{center}
\end{figure}

\begin{figure}[t]
\begin{center}
\includegraphics[width=12.5cm]{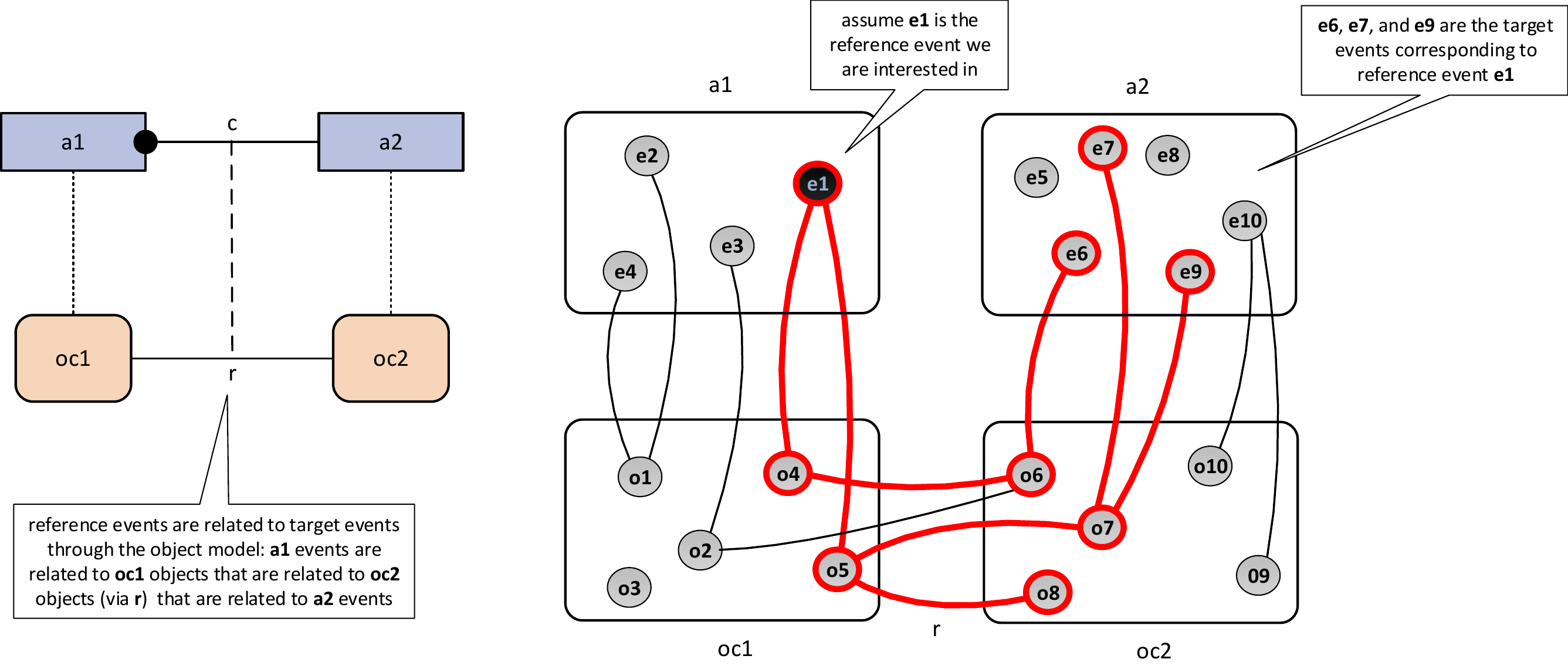}
\caption{Given a reference event for a constraint with
$\mi{crel}(c) = r \in \mi{RT}$ we navigate to the target events through relation $r$ in the object model.}\label{fig-constraint-type2}
\end{center}
\end{figure}

Function $\mi{crel}$ defines the \emph{scope} of each constraint thereby relating reference events to selected target events.
$\mi{crel}(c)$ specifies how events need to be correlated when evaluating constraint $c$. This is needed because we do not assume a fixed case notion
and different entities may interact.
As illustrated by Figure~\ref{fig-two-type-ofconstraints} we basically consider two types of constraints.
In both cases we navigate through the object model to find target events for a given reference event. Figures~\ref{fig-constraint-type1} and \ref{fig-constraint-type2} illustrate how to locate target events.
For each reference event we need the set of all target events in order to check the cardinality constraint.

If $\mi{crel}(c) = \mi{oc} \in \mi{OC}$, then the behavioral constraint is based
on object class $\mi{oc}$.
In Figure~\ref{fig-example1}, $\mi{crel}(\mi{c2}) = \mi{oc2}$.
This means that the target events for constraint $\mi{c2}$ need to be related to the reference events through objects of class $\mi{oc2}$.
Let $e_{\mathit{ref}}$ be the reference event for constraint $\mi{c2}$.
$e_{\mathit{ref}}$ refers to 1 or more $\mi{oc2}$ objects.
The target events of $e_{\mathit{ref}}$ for $\mi{c2}$ are those $\mi{a3}$ events referring to one of these objects.

If $\mi{crel}(c) = r \in \mi{RT}$,
then the target events are related to the reference event through
relations of type $r$ in the object model. Relation $r$ can be traversed in both directions.
In Figure~\ref{fig-example1}, $\mi{crel}(\mi{c1}) = \mi{r1}$
indicating that reference events are related to target events through relationship $\mi{r1}$.
Let $e_{\mathit{ref}}$ be the reference $\mi{a1}$ event for constraint $\mi{c1}$.
$e_{\mathit{ref}}$ refers to $\mi{oc1}$ objects that are related to $\mi{oc2}$ objects through $\mi{r1}$ relations.
The target events of $e_{\mathit{ref}}$ for $\mi{c1}$ are those $\mi{a2}$ events referring to one of these $\mi{oc2}$ objects.

We have now introduced all the modeling elements used in Figure~\ref{fig-intro}.
Note that \emph{create order} activities are related to \emph{pick item} activities
through the relationship connecting class \emph{order} with class \emph{order line}.

\subsection{Discussion}

The graphical notation introduced (e.g., like in Figure~\ref{fig-intro}) fully defines an OCBC model. To illustrate this let us consider a completely different example.
\begin{figure}[tbh]
\begin{center}
\includegraphics[width=12.5cm]{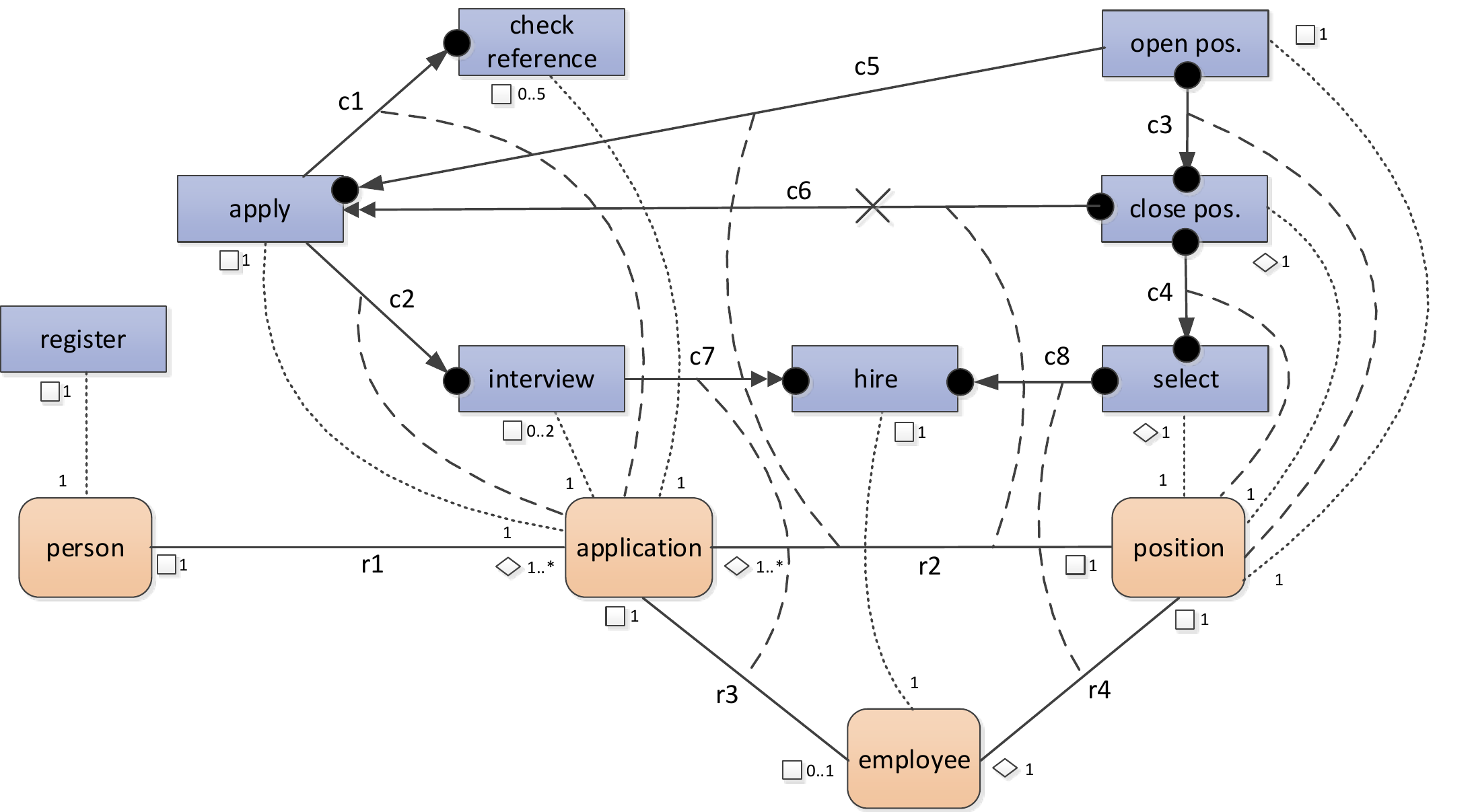}
\caption{An OCBC model modeling a hiring process.}\label{fig-hire-full}
\end{center}
\end{figure}

Figure~\ref{fig-hire-full} models a hiring process.
An organization may create a position. People can apply for such a position, but need to register first. Applications for a position are only considered in the period between opening the position and closing the application process for the position. An application may be followed by at most five reference checks and at most two interviews. In the end one person is selected and subsequently hired for the position.

There are four object classes in the OCBC model:
\emph{person},
\emph{application},
\emph{position},
and \emph{employee}.
The cardinality constraints in Figure~\ref{fig-hire-full}
show that:
each application always refers to precisely one person and one position,
each person eventually applies for some position,
for every position there will eventually be an application,
each employee refers to precisely one application and position,
each application refers to at most one employee, and
each position will eventually refer to one employee.

There is a one-to-one correspondence between registrations (activity \emph{register}) and persons (class \emph{person}).
Activities \emph{apply}, check \emph{reference}, and \emph{interview} each refer
to the class \emph{application}.
Activity \emph{apply} creates one new \emph{application} object.
Activities \emph{open pos.}, check \emph{close pos.}, and \emph{select} each refer to the class \emph{position}.
Activity \emph{open pos.} creates one new \emph{position} object.
There is also a one-to-one correspondence between hirings (activity \emph{hire}) and employees (class \emph{employee}).

Let us now consider the constraints in more detail:
\begin{compactitem}
  \item Constraint $c1$ specifies that every reference check should be preceded by precisely one corresponding application (unary-precedence constraint).
  \item Constraint $c2$ specifies that every interview should be preceded by precisely one corresponding application (unary-precedence).
  \item Constraint $c3$ combines a unary-response and a unary-precedence constraint stating that the opening a a position should be followed by the closing of the application process and the closing should be preceded by the opening of the position.
  \item Constraint $c4$ also combines a unary-response and a unary-precedence constraint stating that the two activities are executed in sequence.
  \item Constraint $c5$ specifies that applications for a position need to be preceded by the opening of that position.
  \item Constraint $c6$ specifies that after closing a position there should not be any new applications for this position (non-response constraint).
  \item Constraint $c7$ specifies that every hire needs to be preceded by at least one interview with the candidate applying for the position (precedence constraint).
  \item Constraint $c8$ again combines a unary-response and a unary-precedence constraint stating that the two activities are executed in sequence.
\end{compactitem}

\begin{figure}[tbh]
\begin{center}
\includegraphics[width=12cm]{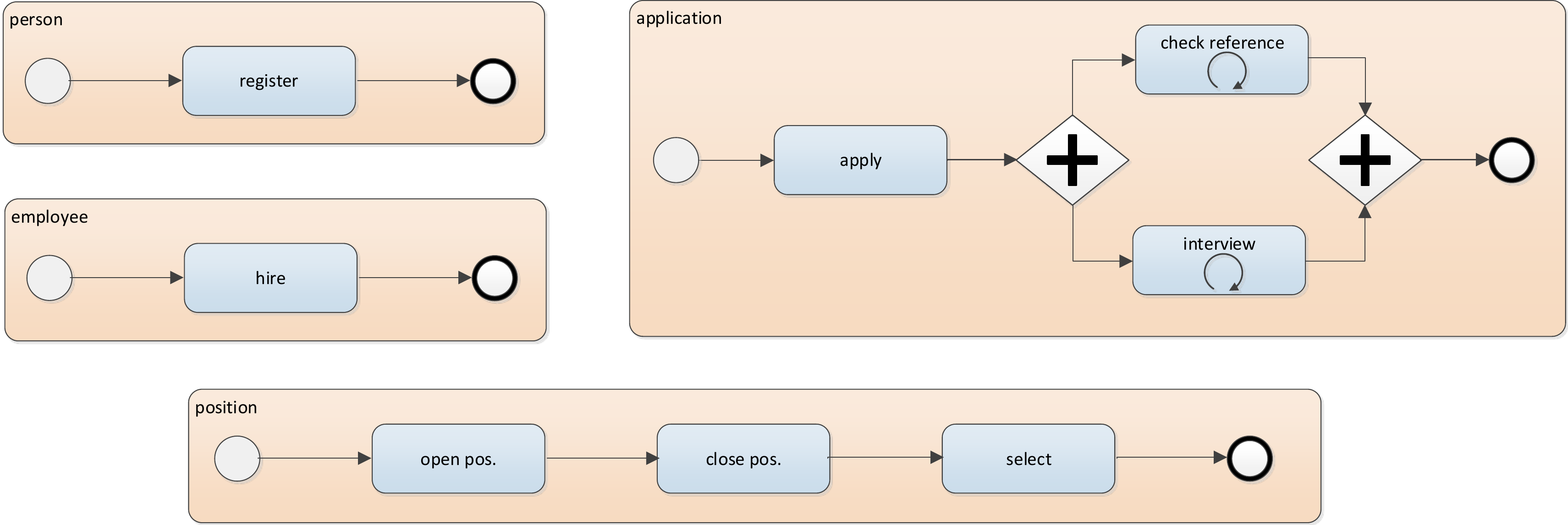}
\caption{An attempt to capture the OCBC model of Figure~\ref{fig-hire-full} in terms of four BPMN models. The relations with the overall data model and interactions between the different entities are no longer visible. For example,
insights like ``one can only apply if the corresponding position is opened but not yet closed'' and ``only people that had an interview can be hired'' get lost.  }\label{fig-BPMN-hire}
\end{center}
\end{figure}

It is important to note that the constraints are based on the object model and that there is not a single instance notion. To illustrate this consider the BPMN model in Figure~\ref{fig-BPMN-hire} which models the lifecycles of
persons, positions, applications, and employees in separate diagrams.
The BPMN model looks very simple, but fails to capture dependencies between the different entities. Consider for example constraints $c5$, $c6$, $c7$, and $c8$ in the OCBC model of Figure~\ref{fig-hire-full}.
The BPMN model does \emph{not} indicate that there is a one to many relationship between positions and applications, and does \emph{not} show that one can only apply if the corresponding position is opened but not yet closed.
The BPMN model does \emph{not} indicate that only one person is hired per position
and that the person to be hired should have registered, applied, and had at least one interview.
The BPMN model does \emph{not} indicate that employees are hired after the completion of the selection process. Note that the same person could apply for multiple positions and many people may apply for the same position.
Obviously this cannot be captured using a single process instance (case) notion.

Comparing Figure~\ref{fig-hire-full} and Figure~\ref{fig-BPMN-hire} reveals
that modeling the lifecycles of entities separately, like in artifact-centric approaches, is not sufficient to capture the real process. The individual lifecycles are simple, but fail to reveal the interplay between
persons, positions, applications, and employees.

\emph{It is essential to understand that the scoping of events considered in a constraint is done through the object model.} This provides a tight integration
between behavior and structure. Moreover, the approach is much more general and more expressive than classical approaches where events are correlated through cases.
Normally, process models (both procedural and declarative) describe the lifecycle of a process instance (i.e., case) in isolation.
This implies that events are partitioned based on case identifiers and different cases cannot share events.
Hence, one-to-many and many-to-many relationships cannot be modeled (without putting instances in separate subprocesses, artifacts or proclets).
In fact, more complicated forms of interaction cannot be handled.
\begin{figure}[tbh]
\begin{center}
\includegraphics[width=8cm]{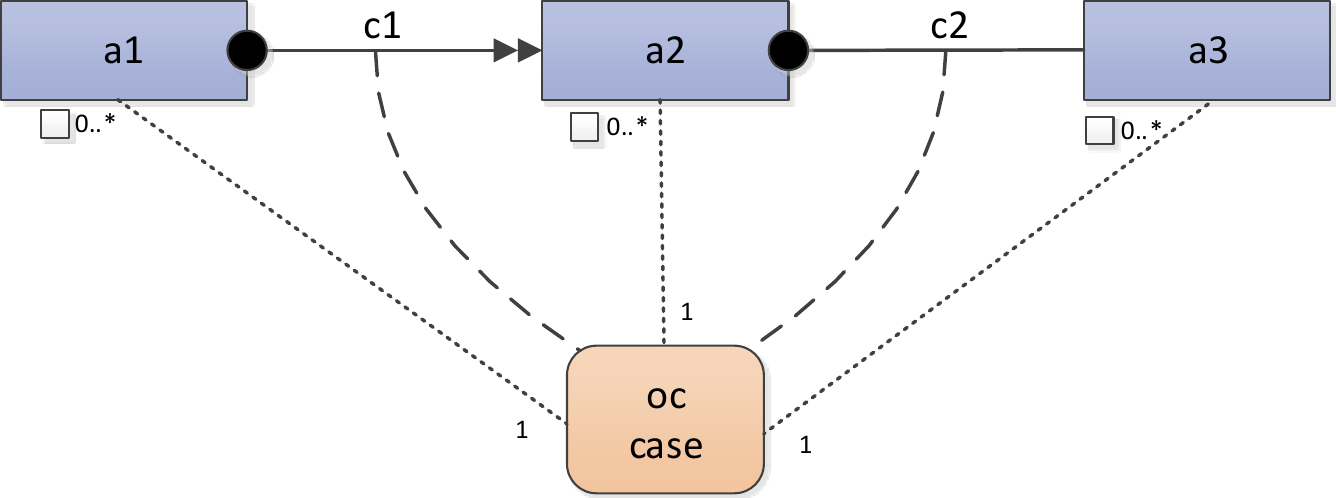}
\caption{An OCBC model mimicking the classical situation where behavior needs to be straightjacketed in isolated process instances (i.e., cases).}\label{fig-example2}
\end{center}
\end{figure}

Note that traditional single-instance modeling approaches can still be mimicked
by using an object model
having one object class $\mi{case}$ and $\mi{crel}(c) = \mi{case}$ for
each constraint $c$. Figure~\ref{fig-example2} sketches this situation
and illustrates that the classical view on process behavior is every
limiting, since complex relationships cannot be captured, and the link to data/object models is missing.

\section{Conformance Checking Using OCBC Models}\label{sec:conf-check}

Given an event log, an object model, and an object-centric behavioral constraint model,
we want to check whether reality (in the form of an event log $L$ and
an object model $\mi{OM}$) conforms to the model $\mi{OCBCM}$.
We identify nine types of possible conformance problems. Most of these problems are not captured by existing conformance checking approaches \cite{wires-replay,BIS-artifactconformance-lnbip2011,BIS-data-aware-checking-lnbip2012,anne_confcheck_is}.

First, we implicitly provide operational semantics for OCBC models by defining a \emph{conformance relation} between event log and model.

\begin{definition}[Conformance]\label{def:confcheck}
Let
$\mi{OCBCM} = (\mi{BCM},\allowbreak \mi{ClaM},\allowbreak \mi{AOC},\allowbreak \acardalways,\allowbreak \acardeven,\ocard,\mi{crel})$ be an OCBC model, with $\mi{BCM} = (A,C,\pi_{\mathit{ref}},\pi_{\mathit{tar}},\mi{type})$ and $\mi{ClaM} = (\mi{OC},\mi{RT},\pi_1,\allowbreak \pi_2,\allowbreak \cardsrcalways,\allowbreak \cardsrceven,\allowbreak \cardtaralways,\allowbreak \cardtareven)$.
Let $L = (E,\mi{act}, \mi{attr}, \mi{EO}, \allowbreak \mi{om}, \preceq)$ be an event log.

Event log $L$ \emph{conforms} to the object-centric behavioral constraint model $\mi{OCBCM}$ if and only if:

\begin{itemize}
  \item \textbf{There are no Type I problems (validity of object models):} for any $e \in E$: object model  $\mi{OM}_{e} = (\mi{Obj}_{e},\mi{Rel}_{e},\mi{class}_{e})$ is valid for $\mi{ClaM}$ (this includes checking the $\alwys$-cardinality constraints that should always hold as stated in Definition~\ref{def:valopmod}),
  \item \textbf{There are no Type II problems (fulfilment):} there is an event $e_f \in \eincafter{e}{E}$ such that for any $e' \in \eincafter{e_f}{E}$:  $\mi{OM}_{e} = (\mi{Obj}_{e},\mi{Rel}_{e},\mi{class}_{e})$ is also fulfilled
      (this involves checking the $\eventy$-cardinality constraints that should eventually hold as stated in Definition~\ref{def:valopmod}),
  \item \textbf{There are no Type III problems (monotonicity):} for any $e_1,e_2 \in E$ such that $e_1 \prec e_2$: $\mi{Obj}_{e_1} \subseteq \mi{Obj}_{e_2}$ and $\mi{class}_{e_1} \subseteq \mi{class}_{e_2}$ (objects do not disappear or change class in-between events).
  \item \textbf{There are no Type IV problems (activity existence):}
  $\{ \mi{act}(e) \mid e \in E\} \subseteq A$  (all activities
    referred to by events exist in the behavioral model),

  \item \textbf{There are no Type V problems (object existence):}
  for all $(e,o) \in \mi{EO}$: $o \in \mi{Obj}_{e}$ (all objects referred to by an event exist in the object model when the event occurs),\footnote{Combined with the earlier requirement, this implies that these objects also exist in later object models.}

  \item \textbf{There are no Type VI problems (proper classes):}
      $\{(\mi{act}(e),\mi{class}_{e}(o)) \mid (e,o) \in \mi{EO}\} \subseteq \mi{AOC}$ (events do not refer to objects of unrelated classes).

  \item \textbf{There are no Type VII problems (right number of events per object):} for any $(a,\mi{oc}) \in \mi{AOC}$, $e \in E$, and $o \in \partial_{\mi{oc}}(\mi{Obj}_e)$:

\begin{itemize}
 \item for any $e' \in \eincafter{e}{E}$:
 $|\{  e'' \in \partial_{a}(\eincbefore{e'}{E}) \mid (e'',o) \in \mi{EO}\}|
 \in \acardalways(a,\mi{oc})$ (each object $o$ of class $\mi{oc}$ has the required number of corresponding $a$ events),

  \item there exists a future event $e_f \in \eincafter{e}{E}$ such that for any $e' \in \eincafter{e_f}{E}$:
 $|\{  e'' \in \partial_{a}(\eincbefore{e'}{E}) \mid (e'',o) \in \mi{EO}\}|
 \in \acardeven(a,\mi{oc})$ (each object $o$ of class $\mi{oc}$ eventually has the required number of corresponding $a$ events),
 \end{itemize}

  \item \textbf{There are no Type VIII problems (right number of objects per event):} for any $(a,\mi{oc}) \in \mi{AOC}$, $e \in \partial_{a}(E)$:
  $|\{  o \in \partial_{\mi{oc}}(\mi{Obj}_e) \mid (e,o) \in \mi{EO}\}|
 \in \ocard(a,\mi{oc})$ (each event $e$ corresponding to activity $a$  has the required number of corresponding objects of class $\mi{oc}$).

 \item \textbf{There are no Type IX problems (behavioral constraints are respected):} for each constraint $c \in C$ and reference event $e_{\mathit{ref}} \in \partial_{\pi_{\mathit{ref}}(c)}(E)$:
 there exists a future event $e_f \in E$ such that for any $e' \in \eincafter{e_f}{E}$:
     $(|\ebefore{e_{\mathit{ref}}}{E_{\mi{tar}}} |,|\eafter{e_{\mathit{ref}}}{E_{\mi{tar}}}|) \in  \mi{type}(c)$ where
  \begin{itemize}
       \item $E_{\mi{tar}} = \{ e_{\mi{tar}} \in \partial_{\pi_{\mathit{tar}}(c)}(E) \mid \exists_{o \in \partial_{\mi{oc}}(\mi{Obj}_{e'})} \ \{(e_{\mathit{ref}},o),(e_{\mi{tar}},o)\} \subseteq \mi{EO}\}$
           if $\mi{crel}(c) = \mi{oc} \in \mi{OC}$,
       \item $E_{\mi{tar}} = \{ e_{\mi{tar}} \in \partial_{\pi_{\mathit{tar}}(c)}(E) \mid \exists_{o_1,o_2 \in \mi{Obj}_{e'}} \  (\{(r,o_1,o_2),(r,o_2,o_1)\} \cap \mi{Rel}_{e'} \neq \emptyset) \ \wedge \  \{(e_{\mathit{ref}},o_1), (e_{\mi{tar}},o_2)\} \subseteq \mi{EO}\}$
           if $\mi{crel}(c) = r \in \mi{RT}$.
   \end{itemize}
 \end{itemize}
\end{definition}

Any event log $L$ that exhibits none of the nine problems mentioned is conforming to $\mi{OCBCM}$. Therefore, one can argue that Definition~\ref{def:confcheck} provides
operational semantics to OCBC models. However, the ultimate goal is not to provide semantics, but to check conformance and provide useful diagnostics.
By checking conformance using Definition~\ref{def:confcheck}, the following four broad classes of problems may be uncovered:
\begin{compactitem}
\item Type I, II, and III problems are related to the object models attached to the events (e.g., object models violating cardinality constraints).
\item Type IV, V, and VI problems are caused by events referring to things that do not exist (e.g., non-existing activities or objects).
\item Type VII and VIII problems refer to violations of cardinality constraints between activities and object classes.
\item Type IX problems refer to violations of the behavioral constraints (e.g., a violation of a response or precedence constraint).
\end{compactitem}

The first two categories (Type I-VI problems) types are more of a bookkeeping nature and relatively easy to understand. The two categories (VII, VIII, IX problems) are related to the more subtle interplay between
activities, objects, relations, and the behavior over time. These are more  interesting, but also quite difficult to understand. Therefore, we elaborate on Type VII, VIII, IX problems.
\begin{figure}[tbh]
\begin{center}
\includegraphics[width=8cm]{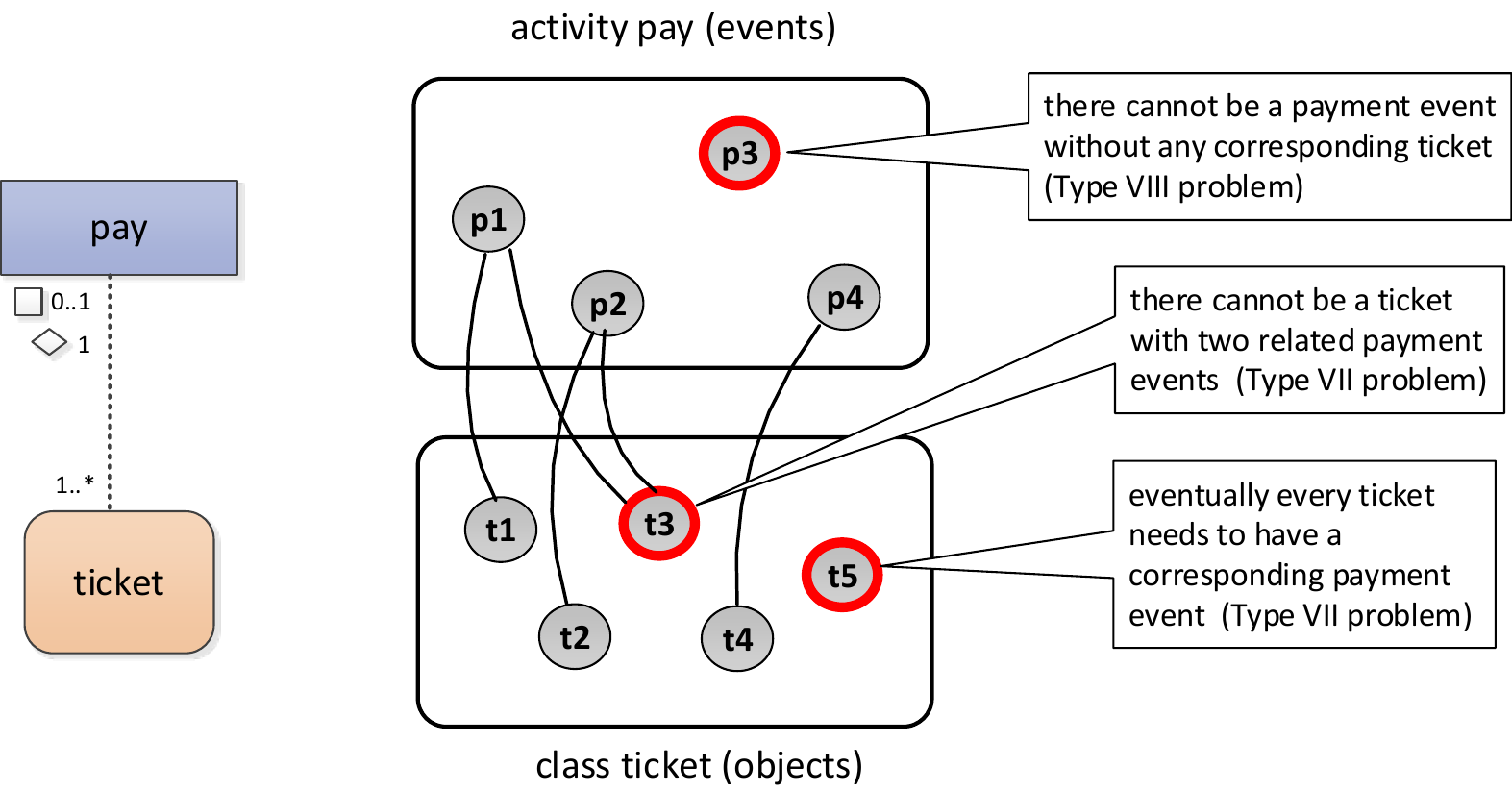}
\caption{An illustration of Type VII and VIII problems (all related to violations of cardinality constraints between activities and object classes).}\label{fig-problem-type78}
\end{center}
\end{figure}

Figure~\ref{fig-problem-type78} shows a situation with problems of Type VII and Type VIII. Object $t3$ has twee corresponding payment events ($p1$ and $p2$), thus violating the
``$\alwys~0..1$'' annotation.
Object $t5$ has no corresponding payment events, thus violating the
``$\eventy~1$'' annotation (assuming there is no corresponding payment in the future).
Event $p3$ has no corresponding payment events, thus violating the ``$1..^\ast$'' annotation. Note that the object model is evolving while the process is executed. This is not shown in Figure~\ref{fig-problem-type78}, i.e., the diagram should be viewed as a snapshot of the process after four payment events.
\begin{figure}[tbh]
\begin{center}
\includegraphics[width=12.5cm]{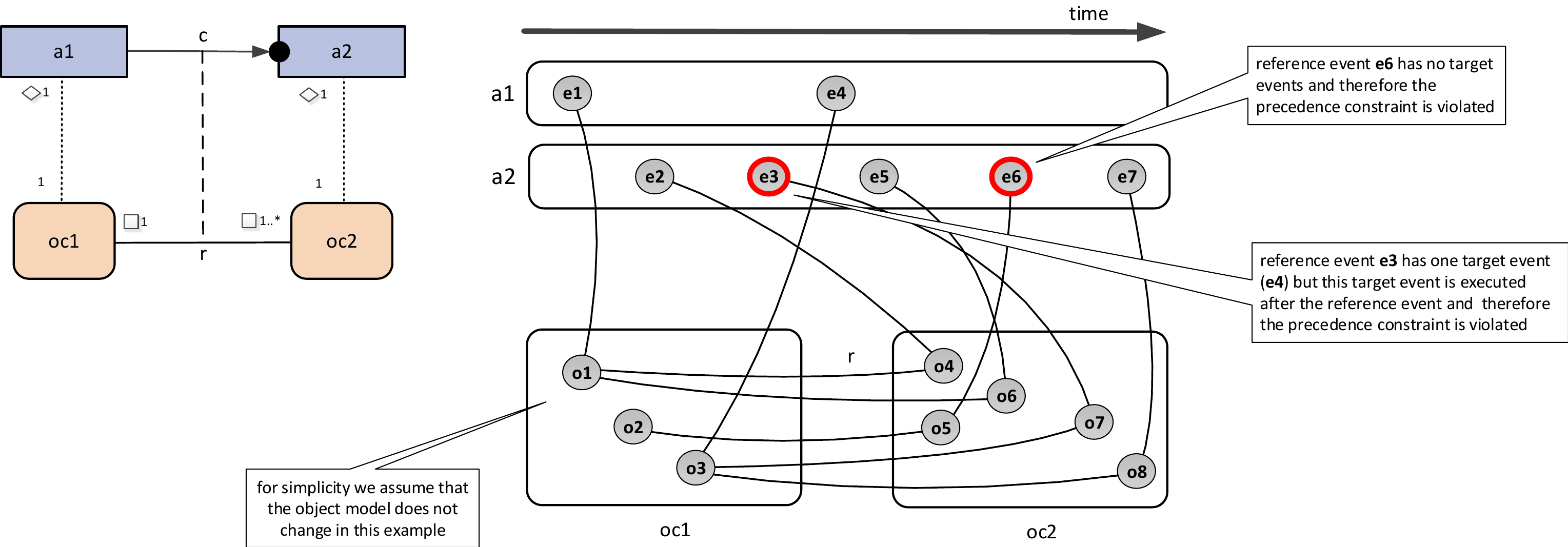}
\caption{An illustration of Type IX problems. The $a1$ and $a2$ events are executed in the order indicated (from left to right). The object model is assumed to remain invariant during the execution of the events (to simplify the explanation). Constraint $c$ is violated for two of the five reference events: both $e3$ and $e6$ have no corresponding $a1$ event that occurred earlier.}\label{fig-problem-type9}
\end{center}
\end{figure}

Figure~\ref{fig-problem-type9} shows a situation with problems of Type IX.
All $a2$ events should have precisely one preceding $a1$ event that is related
through relation $r$. Note that in principle the object model is evolving, but let us assume that all seven events have the object model shown at the lower part of Figure~\ref{fig-problem-type9}.
As stated in Definition~\ref{def:confcheck}, there should be an event $e_f$ after which
the constraint holds for any event $e'$ and corresponding object model $\mi{OM}_{e'}$.
Note that the two cases in the last condition of Definition~\ref{def:confcheck} correspond to the two constraint relations depicted in Figure~\ref{fig-two-type-ofconstraints}.
For each reference event $e_{\mathit{ref}}$, the corresponding set of target events $E_{\mi{tar}}$ is determined by following the links through the object model.
For each $e_{\mathit{ref}}$, the cardinalities are checked: $(|\ebefore{e_{\mathit{ref}}}{E_{\mi{tar}}}|,|\eafter{e_{\mathit{ref}}}{E_{\mi{tar}}}|) \in  \mi{type}(c)$. Hence, it is possible to identify the reference events for which the constraint is violated.

For the situation depicted in Figure~\ref{fig-problem-type9}:
$\mi{type}(c) = \{(\mi{before},\mi{after})\in \Nat \times \Nat \mid \mi{before} = 1\}$,
i.e., there should be precisely one target ($a1$) event preceding 
each reference ($a2$) event related through $r$.
Consider $e2=e_{\mathit{ref}}$ as reference event: $E_{\mi{tar}} = \{e1\}$ and target event $e1$ occurs indeed
before $e2$. Hence, no problem is discovered for $e2$.
Next we consider $e3=e_{\mathit{ref}}$ as reference event: $E_{\mi{tar}} = \{e4\}$, but target event $e4$ occurs \emph{after} $e3$. Hence, $e3$ has no preceding target event signaling a violation of constraint $c$ for reference event $e3$.
If we consider $e5=e_{\mathit{ref}}$ as reference event, we find no problem because $E_{\mi{tar}} = \{e1\}$ and target event $e1$ occurs indeed before $e5$.
If we consider $e6=e_{\mathit{ref}}$ as reference event, we find again a problem because $E_{\mi{tar}} = \emptyset$, so no target event occurs before $e6$.
If we consider $e7=e_{\mathit{ref}}$ as reference event, we find no problem because $E_{\mi{tar}} = \{e4\}$ and target event $e4$ occurs before after $e7$.
Hence, we find two reference event ($e3$ and $e6$) for which constraint $c$ in Figure~\ref{fig-problem-type9} does not hold.

Definition~\ref{def:confcheck} not only provides operational semantics for the graphical notation introduced in this paper,
but also characterizes a wide range of conformance problems.
Following the classification of problems used in Definition~\ref{def:confcheck},
we mention some  possible diagnostics:
\begin{compactenum}
  \item \textbf{Diagnostics for Type I problems (validity of object models):} Highlight
  the $\alwys$-cardinality constraints that make the object model invalid.
  \item \textbf{Diagnostics for Type II problems (fulfilment):} Highlight the $\eventy$-cardinality constraints that do not hold at the end of the log.
  \item \textbf{Diagnostics for Type III problems (monotonicity):} Report objects that disappear or change class over time.
  \item \textbf{Diagnostics for Type IV problems (activity existence):} List the activities appearing in the log and not in the model and highlight the corresponding events in the event log.
  \item \textbf{Diagnostics for Type V problems (object existence):} Highlight all references to non-existing objects.
  \item \textbf{Diagnostics for Type VI problems (proper classes):}  Highlight the events that refer to classes they should not refer to. This can also be shown at the model level, e.g., counting how many times an event corresponding to activity $a$ incorrectly refers to class $oc$.
  \item \textbf{Diagnostics for Type VII problems (right number of events per object):}
  Highlight the $(a,\mi{oc})$ connection if objects in $\mi{oc}$ do not (eventually/always) have the required number of $a$ events. One can count the number of violations and annotate the connections. These violations can also be shown in the event log.
  \item \textbf{Diagnostics for Type VIII problems (right number of objects per event):}
Highlight the $(a,\mi{oc})$ connection if $a$ events do not refer to the specified number of objects in class $\mi{oc}$. One can count the number of violations and annotate the connections. The corresponding events can also be highlighted in the event log.
 \item \textbf{Diagnostics for Type IX problems (behavioral constraints are respected):}
 Highlight the constraints that are violated. Per constraint one can count the number of reference events for which the constraint is violated. These reference events can also be highlighted in the event log.
 \end{compactenum}

The types diagnostics and checks needed are very different from
existing conformance checking approaches.
Most of the conformance checking approaches \cite{wires-replay,BIS-artifactconformance-lnbip2011,anne_confcheck_is} only consider control-flow
and are unable to uncover the above problems. Recently, conformance checking approaches based on alignments have been extended to also check
conformance with respect to the data perspective \cite{BIS-data-aware-checking-lnbip2012}. However, these do not consider a data model
and focus on one instance at a time.

Constraints may be \emph{temporarily violated} while the process is running \cite{declareCSRD09,coloredautomata}.
Consider for example a response constraint involving activities $a$ and $b$: after executing activity $a$ the constraint is
temporarily violated until activity $b$ is executed. This notion exists in any modeling language where process instances need to terminate and is not limited to declarative languages.
Interestingly, the addition of an object may also create temporarily violated and
permanently violated constraints. Consider Figure~\ref{fig-intro} again.
Adding an \emph{order line} object without creating a corresponding \emph{order} results in a permanent violation. However, adding an \emph{order line} while also creating an \emph{order} creates a cascade of obligations:
the obligation to have a \emph{delivery} object, 
the obligation to have a \emph{pick item} event, and
the obligation to have a \emph{wrap item} event.
The corresponding three ``$\eventy~1$'' cardinalities are temporarily violated, 
but can still be satisfied in the future.
Implicitly, there is also the obligation to have a corresponding \emph{deliver items} event in the future.

Interestingly, conformance over OCBC models can
be checked very efficiently. In particular, each of the requirements
in Definition~\ref{def:confcheck} can be formalized as a boolean, SQL-like query over the
input log. The final result is obtained by conjoining all the obtained
answers. This means that the data complexity  of conformance
checking\footnote{That is, the complexity measured in the size of the
  log only, assuming that the OCBC model is fixed.} is in $\textsc{ac}_0$. Recall that
$\textsc{ac}_0$ is strictly contained in $\textsc{LogSpace}$, and
corresponds to the complexity of SQL query answering over a relational
database, measured in the size of the database only.

\section{Conclusion}\label{sec:concl}

In this paper, we proposed \emph{Object-Centric Behavioral Constraint} (OCBC) models
as an integrated approach that merges declarative process modeling and data modeling. 
Cardinality constraints are used to \emph{specify structure and behavior in a single diagram}.
In existing approaches, there is often a complete separation between data/structure (e.g., a class model) and behavior (e.g., BPMN, EPCs, or Petri nets).
In OCBC models, different types of instances can interact in a fine-grained manner and the constraints in the class model guide behavior.

OCBC models are particularly suitable for conformance checking. 
Many deviations can only be detected by considering multiple instances
and constraints in the class model. In this paper, we identified nine types of conformance problems that can be detected using OCBC models.
\begin{SCfigure}
\centering
\includegraphics[width=7cm]{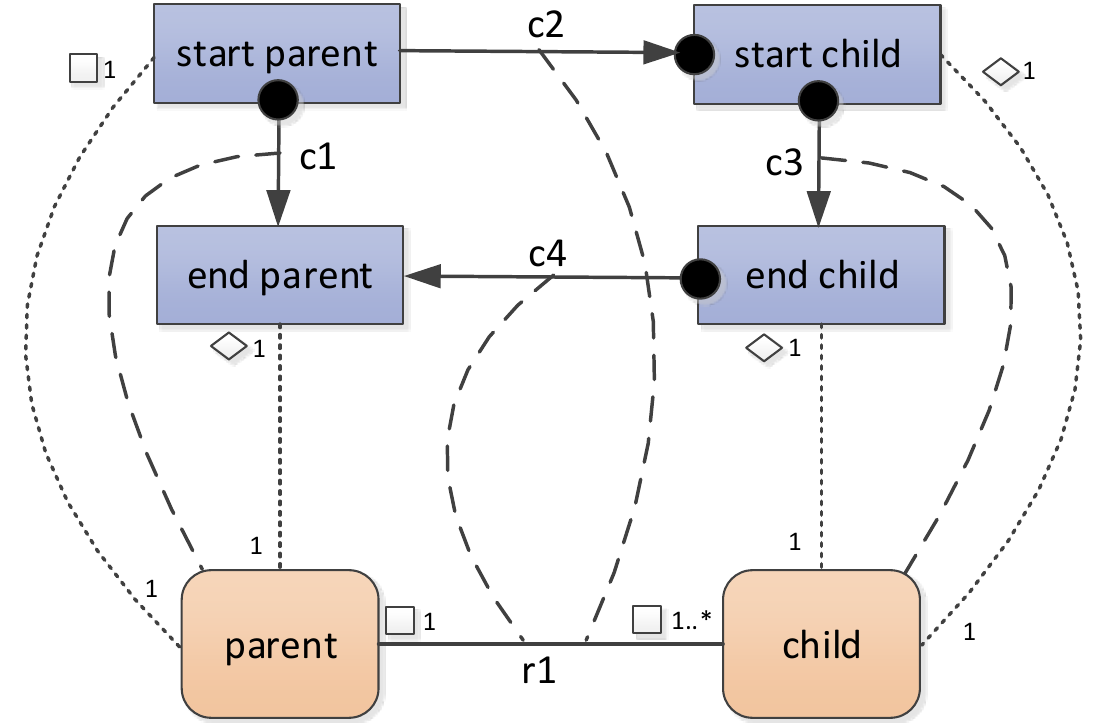}
\caption{Example pattern. After starting the parent, all $k$ children (as defined by $r1$) need to start. After all $k$ children ended, the parent ends. }\label{fig-pattern1}
\end{SCfigure}

The integration of data and control-flow constraints gives raise to sophisticated patterns that cannot be captured in contemporary process modeling approaches. In this light, we  want to identify typical behavioral (anti-)\emph{patterns} that
involve multiple instances or interaction between structure and
behavior.
Figure~\ref{fig-pattern1} shows an example pattern. Along this line, we
plan to study the effect of introducing \emph{subtyping} in the
data model, a constraint present in all data modeling approaches. The
interplay between behavioral constraints and subtyping gives rise to
other interesting behavioral patterns.
For example, \emph{implicit choices} may be introduced through subtyping.
Consider a response constraint pointing to a \emph{payment}
class with two subclasses \emph{credit card payment} and \emph{cash payment}.
Whenever the response constraint is activated and a payment is expected,
such an obligation can be fulfilled by either paying via cash or credit card.

Finally, we also want to investigate how the notions of \emph{consistency} and \emph{constraint conflict/redundancy}, well-known in the
context of Declare \cite{declareCSRD09}, and the corresponding notions of \emph{consistency} and
\emph{class consistency}, well-known in data models \cite{BeCG05},
can be suitably reconstructed and combined in our setting. In this respect, we are currently studying how to formalize OCBC models using temporal description logics, on the one hand to obtain a logic-based semantics for our approach, and on the other hand to derive techniques and decidability/complexity insights on consistency checking and other reasoning tasks.

\bibliographystyle{abbrv}

\bibliography{main-bib}

\end{document}